\documentclass[aps,pra,reprint,amsmath,amsfonts,amssymb,superscriptaddress]{revtex4-1}
\usepackage{graphicx,float,calc}
\usepackage{color,bm}
\usepackage{ulem}
\usepackage{braket}
\usepackage{amsmath}
\usepackage[colorlinks,urlcolor=blue,citecolor=blue,linkcolor=blue]{hyperref}
\begin{document}
	
\title{Identifying non-Hermitian critical points with quantum metric}
	
\author{Jun-Feng Ren}
\affiliation{Key Laboratory of Atomic and Subatomic Structure and Quantum Control (Ministry of Education), Guangdong Basic Research Center of Excellence for Structure and Fundamental Interactions of Matter, South China Normal University, Guangzhou 510006, China}
\affiliation{Guangdong Provincial Key Laboratory of Quantum Engineering and Quantum Materials, School of Physics, South China Normal University, Guangzhou 510006, China}
	
\author{Jing Li}
\affiliation{Key Laboratory of Atomic and Subatomic Structure and Quantum Control (Ministry of Education), Guangdong Basic Research Center of Excellence for Structure and Fundamental Interactions of Matter, South China Normal University, Guangzhou 510006, China}
\affiliation{Guangdong Provincial Key Laboratory of Quantum Engineering and Quantum Materials, School of Physics, South China Normal University, Guangzhou 510006, China}
	
\author{Hai-Tao Ding}
\email{htding@smail.nju.edu.cn}
\affiliation{National Laboratory of Solid State Microstructures and School of Physics, Nanjing University, Nanjing 210093, China}
\affiliation{Collaborative Innovation Center of Advanced Microstructures, Nanjing 210093, China}
\affiliation{Department of Physics, National University of Singapore, Singapore 117551}
	
\author{Dan-Wei Zhang}
\email{danweizhang@m.scnu.edu.cn}
\affiliation{Key Laboratory of Atomic and Subatomic Structure and Quantum Control (Ministry of Education), Guangdong Basic Research Center of Excellence for Structure and Fundamental Interactions of Matter, South China Normal University, Guangzhou 510006, China}
\affiliation{Guangdong Provincial Key Laboratory of Quantum Engineering and Quantum Materials, School of Physics, South China Normal University, Guangzhou 510006, China}

\begin{abstract}
The geometric properties of quantum states is fully encoded by the quantum geometric tensor. The real and imaginary parts of the quantum geometric tensor are the quantum metric and Berry curvature, which characterize the distance and phase difference between two nearby quantum states in Hilbert space, respectively. For conventional Hermitian quantum systems, the quantum metric corresponds to the fidelity susceptibility and has already been used to specify quantum phase transitions from the geometric perspective. In this work, we extend this wisdom to the non-Hermitian systems for revealing non-Hermitian critical points. To be concrete, by employing numerical exact diagonalization and analytical methods, we calculate the quantum metric and corresponding order parameters in various non-Hermitian models, which include two non-Hermitian generalized Aubry-Andr\'{e} models and non-Hermitian cluster and mixed-field Ising models. We demonstrate that the quantum metric of eigenstates in these non-Hermitian models exactly identifies the localization transitions, mobility edges, and many-body quantum phase transitions with gap closings, respectively. We further show that this strategy is robust against the finite-size effect and different boundary conditions.
\end{abstract}
	
\date{\today}
	
\maketitle
	
\section{Introduction}

Quantum phase transitions are a kind of phase transitions that are driven by quantum fluctuations and occur at zero temperature~\cite{Sachdev_2011}. They are extensively investigated between different magnetic phases~\cite{Smacchia2011}, localization phases~\cite{Ganeshan2015,An2021,WangYunfei2022,TangLingZhi2021}, topological phases~\cite{Matsumoto2020,Longhi2019,GongZongping2018,ZhangGuo-Qing2021}, and so on. Near the phase transition point, the infinitesimal variation of certain system parameters will dramatically change the properties of the ground (eigen) state, which intuitively leads to a sharp change of the state fidelity. Moreover, an information-geometric approach based on this observation has already been put forward~\cite{ZanardiPaolo2006,Zhu2006,Zhu2008,zanardi2007information,YouWen-Long2007,GU2010,LvTing2022A,LvTing2022B,ma2010abelian,Carollo2020}. The concrete procedure is to calculate the fidelity susceptibility originally defined in Ref. \cite{YouWen-Long2007}, which characterizes the response of fidelity
to the driving parameter of the system Hamiltonian near critical points.

The fidelity susceptibility is closely related to the quantum metric, which characterizes the distance between two nearby quantum states in Hilbert space~\cite{Michael2017,ding2024non,DingHai-Tao2022,Essay2023}.
The quantum metric is the real part of a more general geometric concept, called the quantum geometric tensor (QGT)~\cite{Michael2017,provost1980}, of which the imaginary part is the well-studied Berry curvature~\cite{Thouless1982,Simon1983,berry1984} for plenty of celebrated physical effects~\cite{GuoWei2016,sundaram1999wave,nagaosa2010anomalous,aharonov1959significance,hasan2010colloquium,qi2011topological,zhang2018topological}.
The quantum metric is also related to many fascinating physical phenomena, such as the superfluidity in flat bands~\cite{julku2016geometric,he2021geometry} and topological quantum phases~\cite{roy2014band,lim2015geometry,palumbo2018revealing}. The quantum metric has recently been experimentally measured in various artificial quantum systems~\cite{TanXinsheng2019,TanXinsheng2021,liao2021experimental,chen2022synthetic,asteria2019measuring,yu2022quantum,zheng2022measuring,lysne2023quantum,tan2018topological,yu2020experimental,yu2022experimental,gianfrate2020measurement,yi2023extracting,cuerda2024observation}.

Most of these previous studies focus on the QGT and related quantum phase transitions in  Hermitian systems \cite{SolnyshkovD.D.2021,CamposVenuti2007,ZanardiPaolo2007,zhang2023,Abasto2008,hetenyi2023fluctuations,Albuquerque2010,Anshuman2012,zhang2023critical}.
In recent years, the theoretical and experimental explorations of non-Hermitian physics have garnered significant attentions~\cite{ashida2020non,HatanoNaomichi1996,HatanoNaomichi1997,BenderCarlM.1998,LiuTao2019,kawabata2019,Chen2018,DWZhang2020,hu2023,Tu2023,TzengYu-Chin2021,Sticlet2023,ora2022,Yao2018,JiangHui2019,Songfei2019,LZTang2020,DWZhang2019,Liu_2020,ZhaiLiang-Jun2020,LiuTong2020,Hamazaki2019,LiJing2023,liao2021experimental,ShanZhong2024}. Various novel phenomena or phases unique to non-Hermitian systems have been revealed, such as exceptional points~\cite{bergholtz2021exceptional,TzengYu-Chin2021,Sticlet2023,ora2022,alvarez2018non,agarwal2022detecting,agarwal2023recognizing}, non-Hermitian skin effects~\cite{Yao2018,Songfei2019,JiangHui2019}, and non-Hermitian delocalization~\cite{LZTang2020,DWZhang2019,Liu_2020,ZhaiLiang-Jun2020,LiuTong2020,Hamazaki2019}. The non-Hermitian quantum metric can characterize the band geometry in pseudo-Hermitian systems that differs from its Hermitian counterpart \cite{ZhuYanQing2021}. It also serves as a tool to detect the symmetry breaking in PT-symmetric systems \cite{DJzhang2019} and becomes a crucial quantity in the dynamics of wave packets near the exceptional points of non-Hermitian systems, diverging in a manner that fully controls the trajectory description of the wave packets \cite{SolnyshkovD.D.2021}. Recent studies have further shown that the quantum metric can characterize topological transitions in the non-Hermitian SSH model \cite{LiJing2023,ChenYe2024}, and the critical scaling near exceptional points can be harnessed to enhance sensing precision \cite{LiJing2023,liao2021experimental,Di2023,Liang2022,HeWan-Ting2023}.

Here we exploit the information-geometric approach to study quantum phase transitions in both single-particle and many-body non-Hermitian systems. We explore the critical points in various non-Hermitian models through the quantum metric. By using the numerical exact diagonalization, we calculate the quantum metric and localization properties (the fractal dimension and participation ratio) in two non-Hermitian generalized Aubry-Andr\'{e} (GAA) models~\cite{harper1955single,aubry1980analyticity,ShanZhong2024}. Our results show that the quantum metric will be divergent or singular in proximity of localization transition points for all eigenstates. Thus the quantum metric exactly identifies the localization transition and the mobility edges in these non-Hermitian quasiperiodic systems. For the non-Hermitian cluster and mixed-field Ising models, we demonstrate that the quantum metric of the ground state can reveal the quantum phase transitions with gap closing, as compared with the corresponding order parameters obtained from both analytical and numerical approaches. Finally, we show that this information-geometric strategy for identifying non-Hermitian critical points is robust against the finite-size effect and different boundary conditions.

The rest of this paper is organized as follows. A brief review of the quantum metric for non-Hermitian Hamiltonians is given in Sec. \ref{sec2}. In Sec. \ref{sec3}, we identify the localization transitions and mobility edges in two non-Hermitian GAA models through the quantum metric of eigenstates, respectively. Sec. \ref{sec4} is devoted to revealing quantum phase transitions in non-Hermitian cluster and mixed-field Ising models with the quantum metric of ground states. Brief discussions on the finite-size effect and boundary conditions and a short conclusion are presented in Sec. \ref{sec5}.

\section{\label{sec2}Non-Hermitian quantum metric}
	
We begin by considering a non-Hermitian Hamiltonian $H(\boldsymbol\lambda)$ parameterized by $\boldsymbol\lambda=(\lambda_1,\lambda_2,\dots,\lambda_D)$ in $D$-dimensional parameter space. The eigenstate equation is $H(\boldsymbol\lambda)\ket{\psi_n^R(\boldsymbol\lambda)}=E_n\ket{\psi_n^R(\boldsymbol\lambda)}$
with $n$ the band index. The eigenvalues $E_n$ are in general complex due to the non-Hermiticity of the Hamiltonian, and the eigenstate $\ket{\psi_1^R(\boldsymbol\lambda)}$ corresponding to the minimum real eigenvalue is set as the ground state. The eigenstates satisfy the normalization condition $\braket{\psi_n^R(\boldsymbol\lambda)|\psi_n^R(\boldsymbol\lambda)}=1$ but is generally nonorthogonal $\braket{\psi_m^R(\boldsymbol\lambda)|\psi_n^R(\boldsymbol\lambda)}\neq0$ $(m\neq n)$. We assume the ground state $\ket{\psi_1^R(\boldsymbol\lambda)}$ is non-degenerate and focus on the right eigenstates $\ket{\psi_n^R(\boldsymbol\lambda)}\doteq\ket{\psi_n(\boldsymbol\lambda)}$ for simplicity. The distance between two neighbor eigenstates $\ket{\psi_n(\boldsymbol\lambda)}$ and $\ket{\psi_n(\boldsymbol\lambda+d\boldsymbol\lambda)}$ in the parameter space is given by~\cite{ye2023quantum,provost1980,Michael2017,Matsumoto2020,sun2022biorthogonal}
\begin{equation}\label{distance}	ds^2=1-\lvert\braket{\psi_n(\boldsymbol\lambda)|\psi_n(\boldsymbol\lambda+d\boldsymbol\lambda)}\rvert^2=\sum_{\mu\nu}g^{(n)}_{\mu\nu}d\boldsymbol\lambda_\mu d\boldsymbol\lambda_\nu,
\end{equation}
where $g^{(n)}_{\mu\nu}$ is the non-Hermitian quantum metric with parameters $\{\mu,\nu\}\in (\lambda_1,\lambda_2,\dots,\lambda_D)$. It is the real part of the non-Hermitian quantum geometric tensor \cite{Matsumoto2020,sun2022biorthogonal}
\begin{equation}\label{nhQM}
\begin{split}
Q^{(n)}_{\mu\nu}&=\braket{\partial_{\boldsymbol\lambda_\mu}\psi_n|\partial_{\boldsymbol\lambda_\nu}\psi_n}-\braket{\partial_{\boldsymbol\lambda_\mu}\psi_n|\psi_n}\braket{\psi_n|\partial_{\boldsymbol\lambda_\nu}\psi_n} \\
&=g_{\mu\nu}-\frac{i}{2}F_{\mu\nu},
\end{split}
\end{equation}
where $g_{\mu\nu}^{(n)}=\text{Re}(Q_{\mu\nu})$, and the imaginary part $F_{\mu\nu}=-2\text{Im}(Q_{\mu\nu})$ corresponds to the Berry curvature.

For identifying the quantum phase transition driven by a single parameter $\mu$, we can consider an infinitesimal change of the parameter: $\mu \to \mu+d\mu$, and then obtain
\begin{equation}\label{exfidelity}	F_n^2\doteq\lvert\braket{\psi_n(\mu)|\psi_n(\mu+d\mu)}\rvert^2\approx1-\chi^{(n)}_F(d\mu)^2.
\end{equation}
Here $F_n$ is the so-called fidelity of the $n$-th eigenstate, and $\chi^{(n)}_F$ is the fidelity susceptibility~\cite{YouWen-Long2007}. The fidelity susceptibility $\chi^{(n)}_F$ denotes the response of $F_n$ due to the small change of $\mu$. For ground states in Hermitian systems, the equivalence of the fidelity susceptibility and the quantum metric element has been shown. For continuous quantum phase transitions with gap closing, the fidelity susceptibility and quantum metric can be used to detect the critical points~\cite{ZanardiPaolo2006,Zhu2006,Zhu2008,zanardi2007information,YouWen-Long2007,LvTing2022A,GU2010,LvTing2022B}. However, this wisdom may fail for discontinuous transitions without gap closing. For instance, it was found that the ground-state fidelity susceptibility is unable to witness the extended-critical phase transition point in the Hermitian p-wave-pair Aubry-Andr\'{e}-Harper model due to the absence of gap closing~\cite{LvTing2022B}.

Combining Eq.~(\ref{distance}) and Eq.~(\ref{exfidelity}) for self-normal right eigenstates, we can find the equivalence of the diagonal element of quantum metric and the fidelity susceptibility in generic non-Hermitian Hamiltonians:
\begin{equation}\label{susceptibility}
\begin{split}
g^{(n)}_{\mu\mu}&=\text{Re}(\braket{\partial_{\mu}\psi_n|\partial_{\mu}\psi_n}-\braket{\partial_{\mu}\psi_n|\psi_n}\braket{\psi_n|\partial_{\mu}\psi_n})
\\&=\lim_{d\mu\to0}\frac{-2\ln F_n}{d\mu^2}=\chi^{(n)}_F.
\end{split}
\end{equation}
For $n$-th eigenstate, one can expect that the corresponding quantum metric $g^{(n)}_{\mu\mu}$ will reach the maximum value or diverge at its localization transition point. In the following, we extend this wisdom to generic non-Hermitian single particle quasi-periodic models and many-body spin models. To be concrete, we demonstrate that the localization transition of all eigenstates with related mobility edges and the quantum phase transitions (with gap closing) in non-Hermitian disordered and many-body systems can be revealed by the quantum metric, respectively.

\section{\label{sec3} Non-hermitian localization transitions and mobility edges}
	
\subsection{Detecting localization transitions}
	
\begin{figure}[tb]
	\centering
	\includegraphics[width=0.48\textwidth]{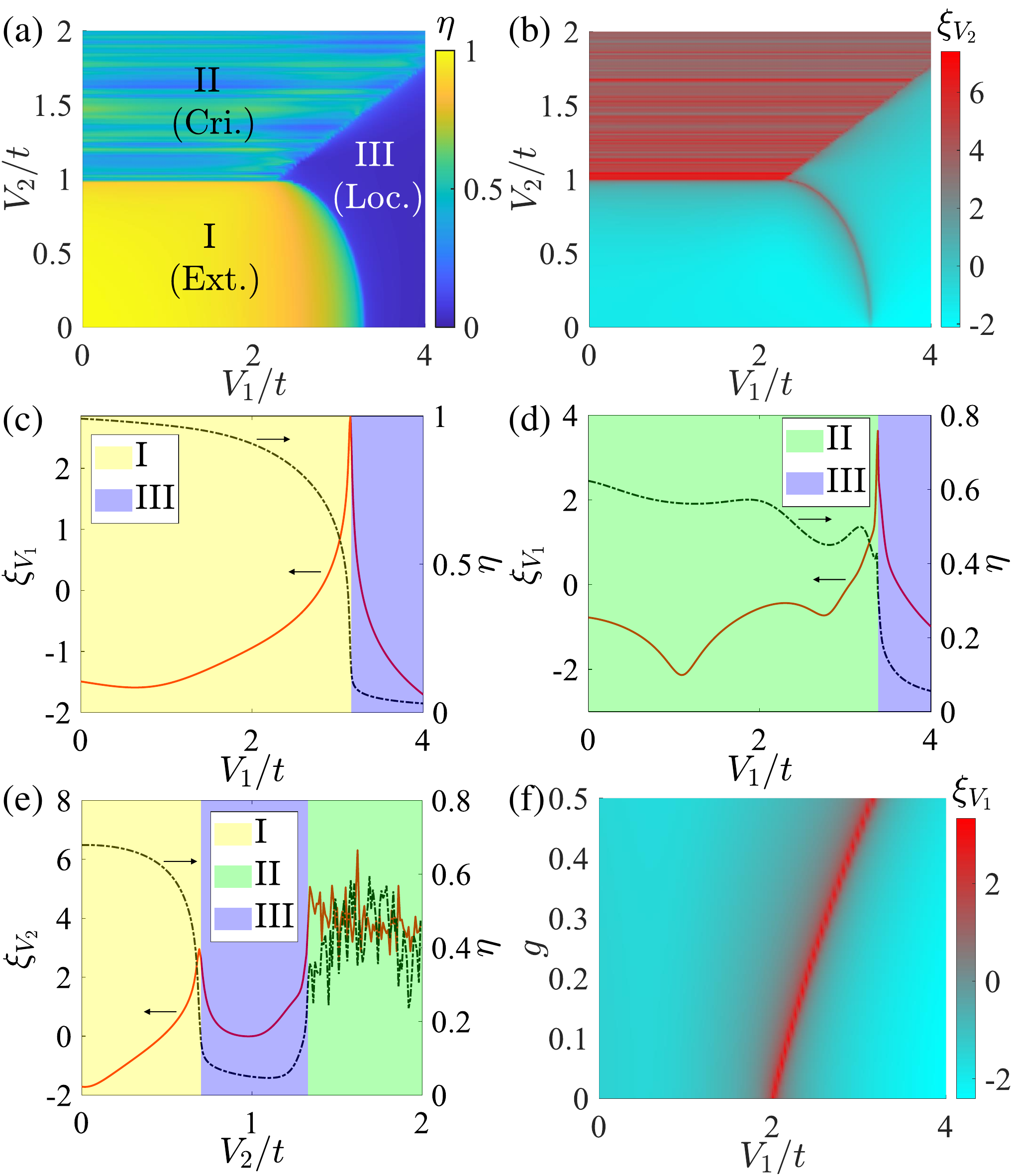}
	\caption{(Color online) The non-Hermitian GAA model with nonreciprocal hopping given by $H_{\text{GAA}}^{(1)}$ with $L=610$ under the PBC. (a) $\eta$ and (b) $\xi_{V_2}$ for the ground state in the parameter space ($V_1$,$V_2$). The localization phase diagram in (a) consists three regions $\mathrm{I}$, $\mathrm{II}$, $\mathrm{III}$ for the extended, critical, and localized phases, respectively. (c,d) $\xi_{V_1}$ and ${\eta}$ as a function of $V_1$ with fixed $V_2=0.5$ and $V_2=1.5$, respectively. (e) $\xi_{V_2}$ and ${\eta}$ as a function of $V_2$ with fixed $V_1=3$. (f) $\xi_{V_1}$ as functions of $V_1$ and $g$ with fixed $V_2=0.5$. The colored phase regions with boundaries in (c-e) are determined by Eq.~(\ref{boundary}). The nonreciprocal strength is $g=0.5$ in (a-e). Other parameters are $h=0$ and $t=1$.
	}\label{fig1}
\end{figure}

\begin{figure}[tb]
	\centering
	\includegraphics[width=0.48\textwidth]{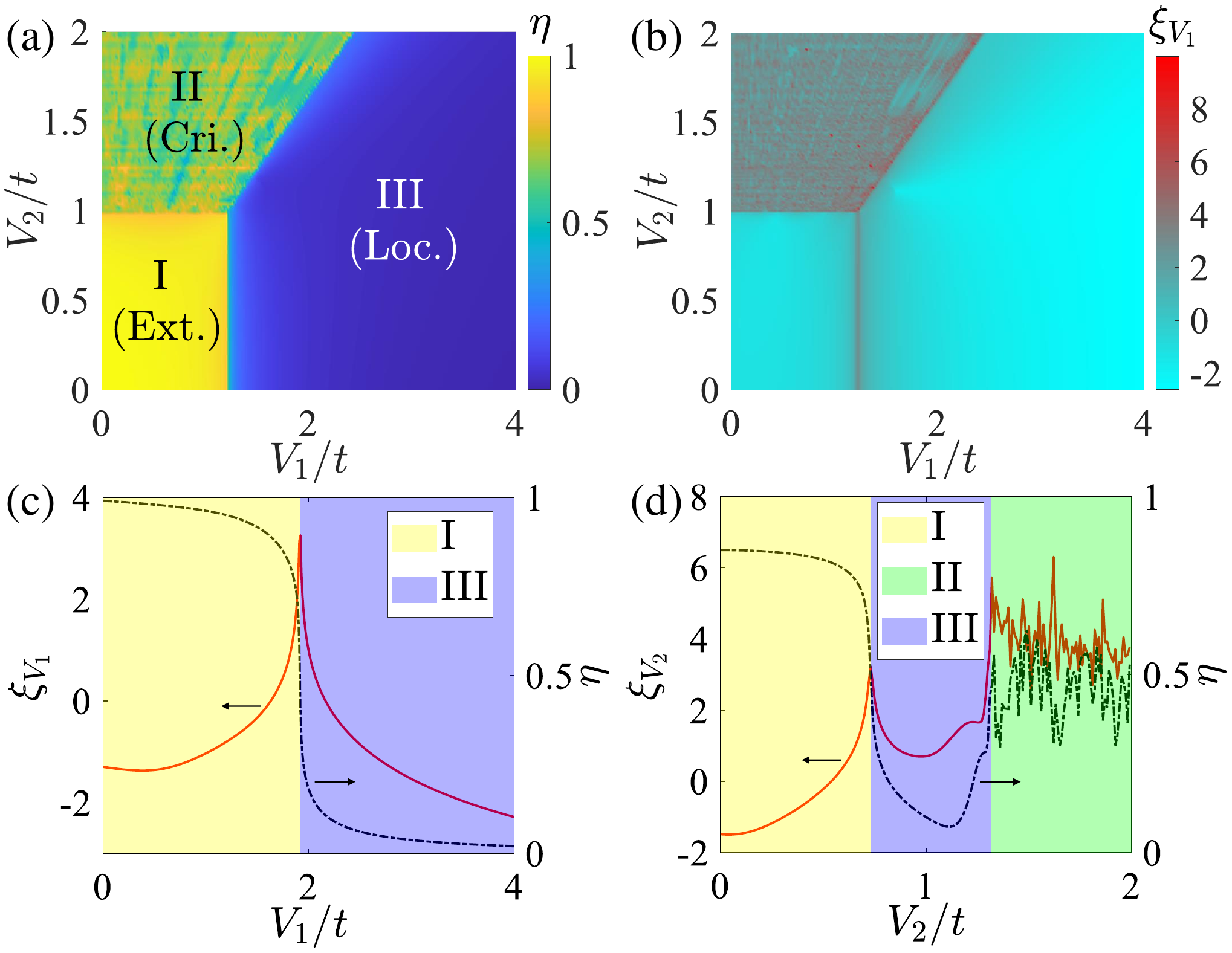}
	\caption{(Color online) The non-Hermitian GAA model with complex potential phase (and nonreciprocal hopping) given by $H_{\text{GAA}}^{(1)}$ with $L=610$ under the PBC. (a) $\eta$ and (b) $\xi_{V_1}$ in the parameter space ($V_1$,$V_2$) for $g=0$. The three regions $\mathrm{I}$, $\mathrm{II}$, and $\mathrm{III}$ denotes the extended, critical, and localized phases, respectively. (c) $\xi_{V_1}$ and ${\eta}$ as a function of $V_1$ with fixed $V_2=0.5$ and $g=0.5$. (d) $\xi_{V_2}$ and ${\eta}$ as a function of $V_2$ with fixed $V_1=1.8$ and $g=0.5$. The colored phase regions with boundaries in (c,d) are determined by Eq.~(\ref{boundary}). Other parameters are $h=0.5$ and $t=1$.
	}\label{fig2}
\end{figure}
	
We first consider a non-Hermitian version of the GAA model with quasi-periodic modulations on both the on-site potential and off-diagonal hopping amplitude \cite{Chang1997,FLiu2015,YWang2021}. Generally, the non-Hermiticities come from the nonreciprocal hopping and complex potential phase, and the model Hamiltonian can be written as~\cite{TangLingZhi2021,harper1955single,aubry1980analyticity}
\begin{equation}\label{AAH}
	H_{\text{GAA}}^{(1)}=\sum_j [t_{j}(e^{-g} c^\dagger_{j+1} c_j+e^{g} c^\dagger_j c_{j+1}) +\Delta_j c^\dagger_j c_{j}],
\end{equation}
where $c_j$ ($c_j^{\dagger}$) is the annihilation (creation) operator at the $j$-th site. Here $g$ is the dimensionless nonreciprocal strength of the hopping amplitude $t_j$, and $\Delta_j$ is the on-site potential term. They are given by
\begin{equation}\label{AAH2}
	\begin{split}
		t_{j}&=t+V_2\cos[2\pi\beta (j+1/2)],\\
		\Delta_j&=V_1\cos(2\pi\beta j+ih),\\
	\end{split}
\end{equation}
with $V_1$ and $V_2$ being the modulation amplitudes of the potential and hopping terms, respectively, $h$ being the complex phase, and $\beta$ being an irrational number to ensure the incommensurate modulation. Without loss of generality, we choose $\beta=(\sqrt{5}-1)/2$ as the golden ratio, which can be obtained from the Fibonacci sequence $F_{j+1}=F_j+F_{j-1}$ as $\beta=\lim_{j\to\infty}F_j/F_{j+1}$. In numerical simulations, we set the lattice site $L=610$ from $F_j$ to satisfy the periodic boundary condition (PBC), which is large enough to detect the critical points for $H_{\text{GAA}}^{(1)}$. This will be clearly seen from the finite-size scaling in Fig. \ref{fig7} in Sec. \ref{sec5}. We set $t=1$ as the unit of energy.

The GAA model has rich localization phase diagrams in the Hermitian \cite{Chang1997,FLiu2015,YWang2021} and non-Hermitian cases \cite{TangLingZhi2021}. By changing the modulation amplitudes $V_1$ and $V_2$ (or $g$), the localization transition of eigenstates between extended, localized, and critical phases can occurs. A typical quantity to characterize localization properties of the $n$-th eigenstate in this model is the fractal dimension:
\begin{equation}\label{fd}
	\eta_n=-\frac{\ln(\sum_j^L|\braket{j|\psi_n}|^4)}{\ln L},
\end{equation}
where $\ket{j}$ is the computational basis and $\ket{\psi_n}$ is the $n$-th right eigenstates. In this model, there is no mobility edge and all eigenstates share the same localization properties \cite{Chang1997,FLiu2015,YWang2021,TangLingZhi2021}. Thus, we focus on the ground state $\ket{\psi_1}$ with the corresponding fractal dimension $\eta_1\equiv\eta$ for simplicity. When the system size $L\to\infty$, one has $\eta\to 1$ for the extended phase, $\eta\to 0$ for the localized phase, while $\eta$ falls between $0$ and $1$ ($\eta\approx0.5$) in the critical phase, respectively. Figure~\ref{fig1}(a) show the numerical results of $\eta$ in the parameter space ($V_1$,$V_2$) for nonreciprocal case with $g=0.5$ and $h=0$ under the PBC, by using the exact diagonalization method. Three regimes (denoted by the I, II and III) for the extended, critical, and localized phases are exhibited. Notably, the critical values $V_{1c}$ for localization transition points with respect to other parameters are given by \cite{TangLingZhi2021}
\begin{equation}\label{boundary}
		V_{1c}=e^{-|h|}\left(2K\cosh|g|+2\sqrt{K^2-V_2^2}\sinh|g|\right),
\end{equation}
with $K = \max(t,V_2)$.
		
To demonstrate the ability of the non-Hermitian quantum metric to detect localization transition points in non-Hermitian cases,we numerically calculate the diagonal elements of the quantum metric $g_{\mu\mu}$ of ground state for the parameter $\mu=\{V_1,V_2\}$, which are equivalent to the ground-state fidelity susceptibility. To resolve the three different localization regions in this model, we use the logarithm quantity of the quantum metric
	\begin{equation}\label{log}
		\xi_{\mu}=\log_{10}g_{\mu\mu},
	\end{equation}
with respect to the model parameters $(V_1,V_2,g,h)$. We first consider the generalized AA model with only nonreciprocal hopping for fixed $h=0$ and $g=0.5$. The numerical result of $\xi_{V_2}$ in Fig.~\ref{fig1}(b) shows the boundaries between different localization phase regions, which is consistent with those in Fig.~\ref{fig1}(a). To reveal the localization transition points more clearly, we plot $\xi_{V_1}$ and $\eta$ as a function of $V_1$ for $V_2=0.5$ and $V_2=1.5$ in Fig.~\ref{fig1}(c) and Fig.~\ref{fig1}(d), respectively. The results show that $\xi_{V_1}$ changes relatively smoothly in each phase, while increases significantly and reaches its maximum value at the localization transition points. In Fig.~\ref{fig1}(e), we plot $\xi_{V_2}$ and $\eta$ as a function of $V_2$ for $V_1=3$, which across three localization phases. In this case, we can see that the boundaries between different phases can also be revealed, although the rapid oscillation of $\xi_{V_2}$ in the critical phase region is exhibited due to the accidental degeneracies of wave functions with respect to $V_2$ in the critical phase. The quantum metric can even be used to identify the localization transition driven by the non-Hermitian strength $g$, as shown in Fig.~\ref{fig1}(f).
		
We also consider the GAA model with the complex potential phase for $h=0.5$ and $g=0$, with numerical results of $\eta$ and $\xi_{V_1}$ shown in Fig.~\ref{fig2}(a) and Fig.~\ref{fig2}(b), respectively. It is evident that the boundaries between the three localization phases distinguished by $\eta$ corresponds to the peak of $\xi_{V_1}$. For the case of coexistence of the nonreciprocal hopping and complex potential phase with $g=h=0.5$, one can still find the peaks of the quantum metric ($\xi_{V_1}$ and $\xi_{V_2}$) at the localization transition points in Fig.~\ref{fig2}(c) and Fig.~\ref{fig2}(d). Moreover, we plot three phase regions with different colors in Figs.~\ref{fig1}(c-e) and Figs.~\ref{fig2}(c,d) according to Eq. (\ref{boundary}). One can see that localization transition points in this non-Hermitian GAA model revealed by the quantum metric are perfectly consistent with those obtained from the fractal dimension and the analytical result \cite{TangLingZhi2021}.

\subsection{Revealing mobility edges}
	
\begin{figure}[tb]
	\centering
	\includegraphics[width=0.48\textwidth]{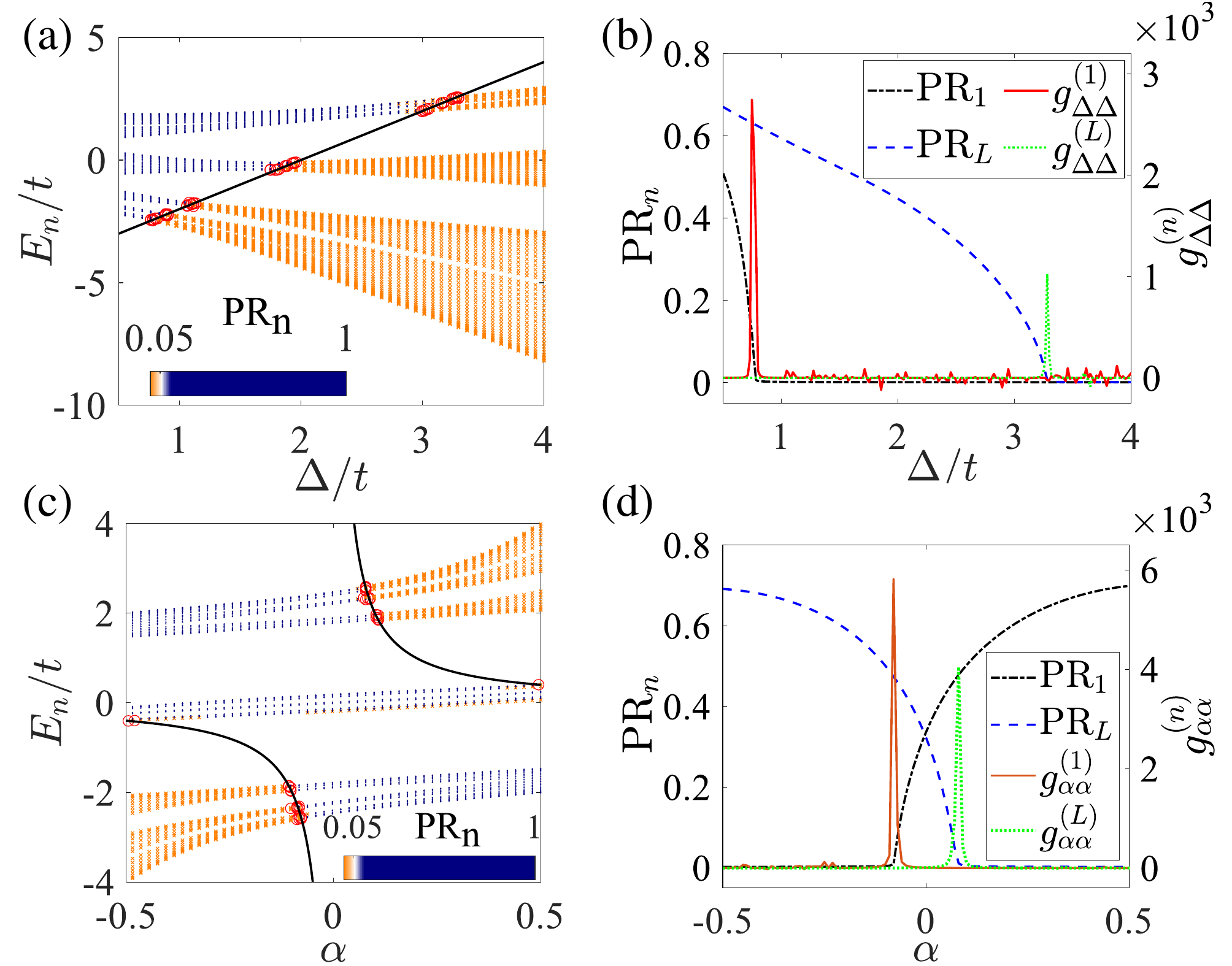}
	\caption{(Color online) The GAA model given by $H_{\text{GAA}}^{(2)}$ in the Hermitian limit with $L=610$ under the PBC. The eigenenergies $E_n$ as functions of (a) $\Delta$ and (c) $\alpha$, respectively. The colors correspond to the values of $\mathrm{PR}_n$ as the inset color bar. The red circles denote the positions of the peaks of the quantum metric for eigenstates at different $\Delta$ or $\alpha$. The black solid line denotes the exact mobility edge given by Eq~(\ref{ME}). Note that $\mathrm{PR}_n=0.05$ is set as the critical value for the comparison with the exact mobility edge. (b) $\mathrm{PR}_n$ and $g_{\Delta\Delta}^{(n)}$ as functions of $\Delta$ for the ground state ($n=1$) and the highest excited state ($n=L$) with fixed $\alpha=-0.5$. (d) $\mathrm{PR}_n$ and $g_{\alpha\alpha}^{(n)}$ as functions of $\alpha$ for the ground state and the highest excited state with fixed $\Delta=1.8$. Other parameters are $g=0$ and $t=1$.
	}\label{fig3}
\end{figure}
	
\begin{figure}[tb]
	\centering
	\includegraphics[width=0.48\textwidth]{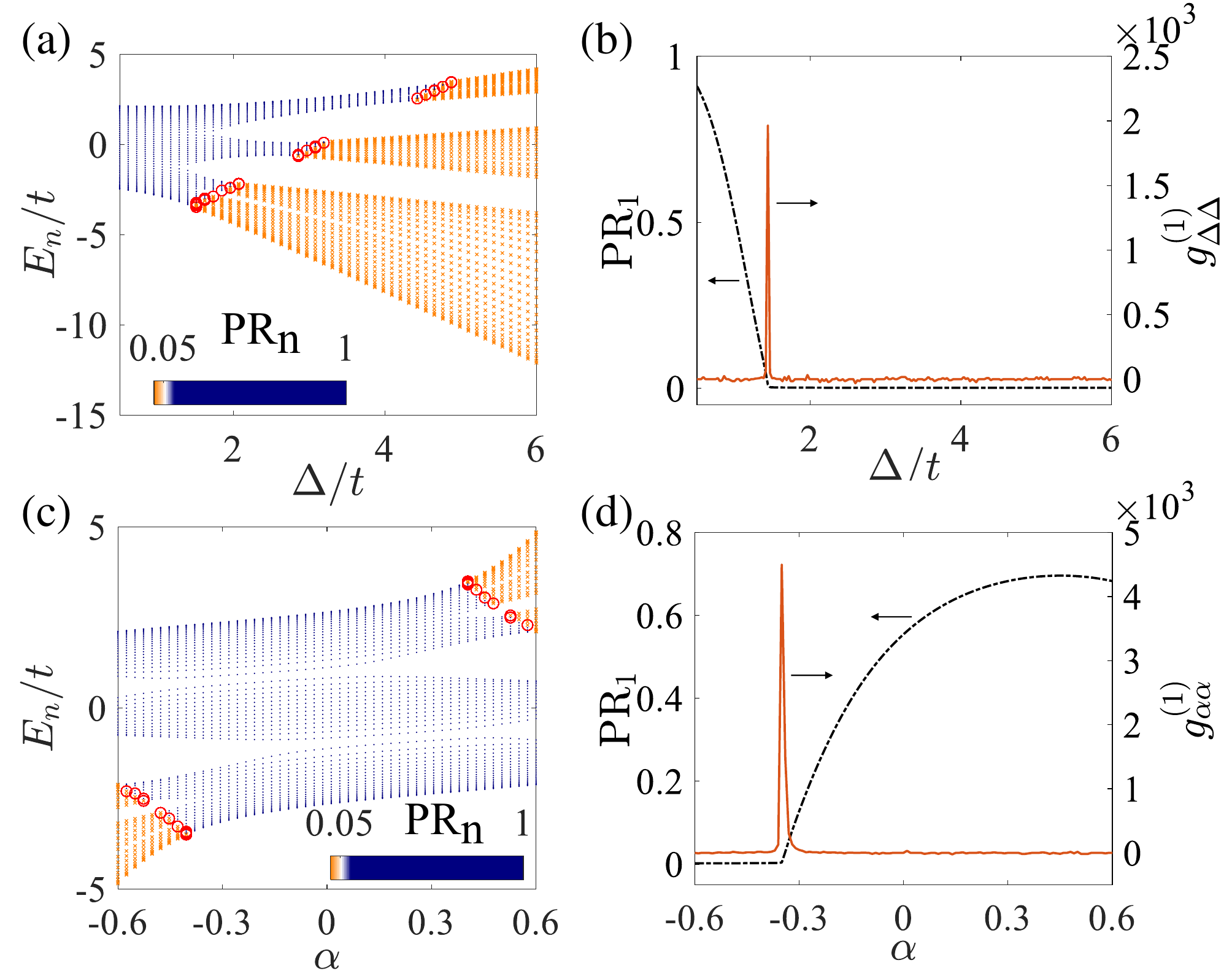}
	\caption{(Color online) The non-Hermitian GAA model given by $H_{\text{GAA}}^{(2)}$ with $L=610$ under the PBC. The eigenenergies $E_n$ as functions of (a) $\Delta$ and (c) $\alpha$, respectively. The red circles denote the positions of the peaks of the quantum metric for eigenstates at different $\Delta$ or $\alpha$. (b) $\mathrm{PR}_1$ and $g_{\Delta\Delta}^{(1)}$ as a function of $\Delta$ for the ground state with fixed $\alpha=-0.5$. (d) $\mathrm{PR}_1$ and $g_{\alpha\alpha}^{(1)}$ as a function of $\alpha$ for the ground state with fixed $\Delta=1.8$. Other parameters are $g=0.5$ and $t=1$.
	}\label{fig4}
\end{figure}
	
We proceed to consider localization transitions of all eigenstates to reveal the mobility edge in non-Hermitian systems with the quantum metric. To do this, we study another non-Hermitian version of the GAA model~\cite{Ganeshan2015,An2021,WangYunfei2022}:
\begin{equation}\label{GAA}
	H_{\text{GAA}}^{(2)}=\sum_j [t(e^{-g} c^\dagger_{j+1} c_j+e^{g} c^\dagger_j c_{j+1}) +\epsilon_j c^\dagger_j c_{j}],
\end{equation}
with a quasi-periodic onsite potential
\begin{equation}\label{GAA2}
	\epsilon_j=\frac{\Delta\cos(2\pi j\beta)}{1-\alpha\cos(2\pi j\beta)}.
\end{equation}
Here $g$ represents the nonreciprocal strength of the hopping amplitude, $\Delta$ is the amplitude of irrational modulation with $\beta=(\sqrt{5}-1)/2$, and $\alpha\in(-1,1)$ denotes the lattice parameter. Here we also choose the lattice size $L=610$ from the Fibonacci sequence to satisfy the PBC. It is sufficiently large for $H_{\text{GAA}}^{(2)}$ and the finite-size scaling will be given in Sec. \ref{sec5}.

In the Hermitian case ($g=0$), when $\alpha=0$, it reduces to the standard AA model without mobility edges. When $\alpha\neq0$, this GAA model in the Hermitian limit has an exact mobility edge at energy $E=E_c$ that follows the relationship~\cite{Ganeshan2015}
\begin{equation}\label{ME}
	E_c=\frac{2t-\Delta}{\alpha}.
\end{equation}
The presence of the mobility edge can be understood by the energy-independent localization transition and self-duality condition driven by the cosine dispersion~\cite{Ganeshan2015,SZLi2024}. This GAA model has been realized with ultracold atoms in tunable synthetic lattices and the mobility edge has been observed \cite{An2021,WangYunfei2022}. However, the mobility edge in the non-Hermitian GAA model is yet to be explored. To characterize the localization properties of all eigenstates and reveal the mobility edge in the Hermitian and non-Hermitian cases, we can numerically compute the participation ratio of the $n$-th eigenstate $|\psi_n\rangle$:
\begin{equation}\label{pr}
	\mathrm{PR}_n=\frac{1}{L}\frac{1}{\sum_j^L|\braket{j|\psi_n}|^4}.
\end{equation}
The extended and localized eigenstates take $\mathrm{PR}_n\to1$ and $\mathrm{PR}_n\to0$, respectively.

To detect the mobility edge with the quantum metric, we numerically calculate the diagonal elements of the quantum metric $g^{(n)}_{\mu\mu}$ with $\mu=\{\Delta,\alpha\}$ for the $n$-th eigenstate $|\psi_n\rangle$, which are equivalent to the fidelity susceptibility
of $|\psi_n\rangle$ with respect to the parameter $\mu$. We first show the numerical results in the Hermitian limit with $g=0$ in Fig.~\ref{fig3}. As shown in Fig.~\ref{fig3}(a), the mobility edge exhibits and separates the eigenstates from $\mathrm{PR}_n$: the upper ones as extended states while the lower ones are localized states. The peak positions of $g^{(n)}_{\Delta\Delta}$ and the exact mobility edge given by Eq.~(\ref{ME}) are plotted, which are consistent with each other. To be more clearly, we show $g^{(n)}_{\Delta\Delta}$ and $\mathrm{PR}_n$ as a function of $\Delta$ for the ground state ($n=1$) and the highest excited state ($n=L$) in Fig.~\ref{fig3}(b), where $g^{(n)}_{\Delta\Delta}$ exhibits a distinct peak at the critical point with $\mathrm{PR}_n\approx0$. The mobility edge separating extended and localized eigenstates with respect to varying the parameter $\alpha$ can be also revealed by the quantum metric $g^{(n)}_{\alpha\alpha}$, as shown in Figs. \ref{fig3}(c,d).

For the non-Hermitian case, the nonreciprocal strength $g$ modify the mobility edge from the exact form, with two examples of $g=0.5$ shown in Fig.~\ref{fig4}. In these cases, the localization transitions of right eigenstates with respect to the parameters $\Delta$ ($\alpha$) are revealed by $g^{(n)}_{\Delta\Delta}$ ($g^{(n)}_{\alpha\alpha}$) in Fig.~\ref{fig4}(a) [\ref{fig4}(c)]. Figure ~\ref{fig4}(b) [\ref{fig4}(d)] shows $\mathrm{PR}_1$ and $g^{(1)}_{\Delta\Delta}$ ($g^{(1)}_{\alpha\alpha}$) as a function of $\Delta$ ($\alpha$) for the ground state. One can find the peak of $g^{(1)}_{\Delta\Delta}$ ($g^{(1)}_{\alpha\alpha}$) is exhibited at the localization transition point. Comparing with the Hermitian case, the mobility edge with respect to $\Delta$ moves to the larger values of $\Delta$ in Fig~\ref{fig4}(a), while it moves towards larger $|\alpha|$ in Figs~\ref{fig4}(c). This can be understood by the delocalization enhanced by the nonreciprocal hopping \cite{HatanoNaomichi1996,HatanoNaomichi1997,GongZongping2018}.
Thus, our results demonstrate the quantum metric can be used to reveal the mobility edge in both the Hermitian and non-Hermitian systems.

\section{\label{sec4} Quantum phase transitions in non-hermitian many-body systems}

\begin{figure*}[tb]
	\centering
	\includegraphics[width=0.98\textwidth]{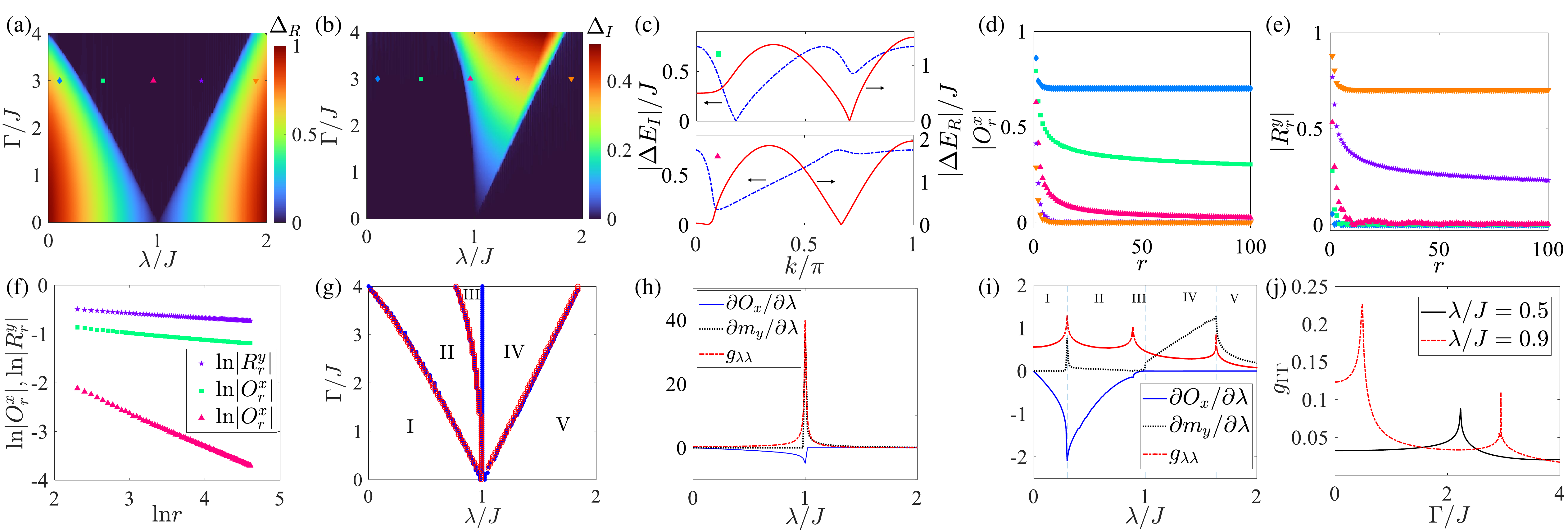}
	\caption{(Color online)The non-Hermitian cluster Ising model $H_{\text{NH-Ising}}^{(1)}$ under the PBC. (a)$\Delta_R$ and (b) $\Delta_I$ as functions of $\lambda$ and $\Gamma$, respectively. The parameters ($\lambda$, $\Gamma$) for the markers in (a-f) are (0.1,3) for the blue diamond, (0.5,3) for the green square, (0.96,3) for the pink up-triangle, (1.4,3) for the purple pentagram, and (1.9,3) for the orange inverted triangle, respectively. (c) $|\Delta E_R|$ and $|\Delta E_I|$ versus $k$ for $\lambda=0.5$ (upper) and $\lambda=0.96$ (lower) with fixed $\Gamma=3$. (d) $|O_r^x|$ and (e) $|R_r^y|$ as a function of $r$, respectively. (f) $\mathrm{ln}|O_r^x|$ and $\mathrm{ln}|R_r^y|$ versus $\mathrm{ln}r$ with the power-law decay behaviour. (g) The phase diagram with boundaries determined from the order parameters (blue solid points) and the quantum metric (red circles) of the ground state, respectively. The regions from I to V correspond to five phases: The real gapped cluster phase in region I, the gapless quasi cluster phase in region II, the imaginary gapped quasi cluster phase in region III, the imaginary gapped antiferromagnetic phase in region $\mathrm{IV}$, and the real gapped antiferromagnetic phase in region $\mathrm{V}$, respectively. $g_{\lambda\lambda}$, $\partial O_x/\partial\lambda$, and $\partial m_y/\partial\lambda$ as a function of $\lambda$ for (h) $\Gamma=0$ and (i) $\Gamma=3$, respectively. (j) $g_{\Gamma\Gamma}$ as a function of $\Gamma$ with $\lambda=0.5,0.9$. Other parameter is $J=1$.
	}\label{fig5}
\end{figure*}

In this section, we proceed to reveal the quantum phase transition of many-body ground states in non-Hermitian interacting spin systems with the diagonal elements of quantum metric, which correspond to the ground-state fidelity susceptibility. Although most of experimental studies of non-Hermitian physics focus on single-particle systems, the extension to non-Hermitian many-body systems has attracted increasing attentions. For instance, the experimental scheme to realize the non-Hermitian mixed-filed Ising model by using Floquet Rydberg atomic arrays has been proposed ~\cite{ruizhe2023}. The cluster-Ising interaction and controllable dispassion can be experimentally achieved in Floquet superconducting qubit arrays~\cite{zhang2022digital}.

\subsection{Non-Hermitian cluster Ising model}
We first consider the non-Hermitian cluster Ising model with exactly solvable ground states~\cite{GuoZheng-Xin2022}, which takes the following Hamiltonian
\begin{equation}\label{cising}
	H_{\text{NH-Ising}}^{(1)}=-J\sum_{l=1}^N \sigma_{l-1}^x \sigma_{l}^z \sigma_{l+1}^x +\lambda\sum_{l=1}^N \sigma_{l}^y \sigma_{l+1}^y +\frac{i\Gamma}{2}\sum_{l=1}^N \sigma_{l}^u,
\end{equation}
where $\sigma_l^{x,y,z}$ denote the Pauli matrices of the $l$-th spin, and the gain and loss is given by $\sigma_l^u=\begin{pmatrix}1&0\\0&0\end{pmatrix}$. There are three parameters in this model: $J$ represents the cluster exchange strength, $\lambda$ denotes the Ising exchange strength, and $\Gamma$ signifies the strength of gain or loss. In the rest of this section, we set $J=1$ as the energy unit.
	
Using the standard Jordan-Wigner transformation and the Fourier transformation under the PBC, we can obtain the Hamiltonian in momentum space
\begin{equation}\label{cisingk}	
H_k=2\sum_{k=0}^{\pi}[iy_k(c_k^{\dagger}c_{-k}^{\dagger}+c_kc_{-k})+z_k(c_k^{\dagger}c_{k}+c_{-k}^{\dagger}c_{-k}-1)],
\end{equation}
where $y_k=\sin(2k)+\lambda\sin(k)$ is real and $z_k=\cos(2k)-\lambda\cos(k)-\frac{i\Gamma}{4}$ is generally complex.
The Hamiltonian in Eq. (\ref{cisingk}) can be rewritten in the terms of the Bogoliubov-de-Gennes formalism
\begin{equation}\label{cisingbdg}		
H_k=\sum_{k=0}^{\pi}
\begin{pmatrix}
	c_k^\dagger&c_{-k}
\end{pmatrix}
\begin{pmatrix}
	z_k&iy_k
	\\-iy_k&-z_k
\end{pmatrix}
\begin{pmatrix}
	c_k\\c_{-k}^\dagger
\end{pmatrix}.
\end{equation}
The eigen-energies are given by $E_{k,\pm}=\pm\sqrt{z_k^2+y_k^2}$ as a function of momentum $k$, where $E_{k,-}$ and $E_{k,+}$ correspond to the ground-state and excitation spectra, respectively.
In order to diagonalize $H_k$, one can take the non-Hermitian Bogoliubov transformation~\cite{TonyE2014},
\begin{equation}\label{nhbb}
	\begin{split}		
		\eta_k=u_k c_k + v_k c_{-k}^\dagger,~~ \eta_{-k}=-v_k c_{k}^\dagger + u_k c_{-k},\\
		\bar{\eta}_k=u_k c_k^\dagger + v_k c_{-k},~~ \bar{\eta}_{-k}=-v_k c_{k} + u_k c_{-k}^\dagger,
	\end{split}
\end{equation}
where the complex factors $u_k$ and $v_k$ are given by
\begin{equation}\label{cisinggs}
	\begin{split}		
		u_k=\frac{z_k+E_{k,-}}{\mathcal{C}},~~ v_k=-\frac{y_k}{\mathcal{C}},
	\end{split}
\end{equation}
with $\mathcal{C}$ being the constant to satisfy $u_k^2+v_k^2=1$. By diagonalizing $H_k$, we obtain the ground state of $H_{\text{NH-Ising}}^{(1)}$ as
\begin{equation}\label{cisinggs}		
	\ket{G}=\frac{1}{\mathcal{N}}\prod_{k=0}^{\pi}[u_k-v_k c_k^{\dagger}c_{-k}^{\dagger}]\ket{\mathrm{0}},
\end{equation}
where $\mathcal{N}=\prod_{k=0}^{\pi}(|u_k|^2+|v_k|^2)$ is the normalization constant, and $\ket{\mathrm{0}}$ denotes the vacuum state without fermion occupied and is defined via $c_k\ket{\mathrm{0}}=0$.

The properties of the ground state can be characterized by the staggered magnetization $m_y$ and the string order parameter $O_x$~\cite{Smacchia2011,GuoZheng-Xin2022}. The staggered magnetization is $m_{y}^2=\lim_{r\to\infty}(-1)^r R_r^y$, where the two-spin correlation $R_r^y$ is given by~\cite{Smacchia2011}
\begin{equation}\label{my}
	\begin{split}
		R_r^y=\braket{\sigma_{l}^y\sigma_{l+r}^y}
		=(-1)^r\braket{A_l B_{l+1}A_{l+1}\dots A_{l+r-1}B_{l+r}}.
	\end{split}
\end{equation}
Here $A_l=c_l^\dagger+c_l$, $B_l=c_l^\dagger-c_l$, $r$ is the relative distance between two spins, and $l$ ($l\in[1,N]$) is an irrelevant index under the PBC.
Using the Wick theorem~\cite{WickG.C.1950}, $R_{r}^{y}$ can be transformed to the Pfaffian of a $2r\times 2r$ skew-symmetric matrix~\cite{TonyE2014}
\begin{equation}\label{ry}
	R^y_r=(-1)^r \rm{pf}\begin{vmatrix}0&-G_{-1}&Q_1&-G_{-2}&Q_2&\dots&-G_{-r}\\
		 &0&G_{0}&S_1&G_{1}&\dots&S_{r-1}\\
		  &&0&-G_{-1}&Q_1&\dots&-G_{-r+1}\\
		  &&&\ddots&\ddots&&\vdots&\\
		  &&&&0&G_0&S_{1}\\
		  &&&&&0&-G_{-1}\\
		  &&&&&&0\end{vmatrix},
\end{equation}
where $\rm{pf}(D)^2 =\rm{det}(D)$ and its positive value is chosen in our calculation. The matrix elements $G_r=\braket{B_l A_{l+r}}=-\braket{A_{l+r} B_{l}}$, $S_r=\braket{B_l B_{l+r}}$, and $Q_r=\braket{A_l A_{l+r}}$ can be derived based on $u_k$ and $v_k$. By taking similar calculations as those given the appendices in Refs.~\cite{TonyE2014,Zhaozhuan2022}, one obtains the expressions of the matrix elements
\begin{equation}\label{Gr}
	\begin{split}					
		G_r=&-\frac{1}{\pi}\int_{0}^{\pi}dk\cos(kr)(\frac{|u_k|^2-|v_k|^2}{|u_k|^2+|v_k|^2})\\&
		+\frac{1}{\pi}\int_{0}^{\pi}dk\sin(kr)(\frac{u_k v_k^*+u_k^* v_k}{|u_k|^2+|v_k|^2}),\\
	\end{split}
\end{equation}
and
\begin{equation}\label{Qr}
	\begin{split}					
		S_r=Q_r=\frac{1}{\pi}\int_{0}^{\pi}dk\sin(kr)(\frac{u_k v_k^*-u_k^* v_k}{|u_k|^2+|v_k|^2})~ (r\neq0),
	\end{split}
\end{equation}
respectively. The second order parameter is the string order $O_x=\lim_{r\to\infty}(-1)^r O^x_r$, where the string correlation is given by~\cite{Smacchia2011}
\begin{equation}\label{oxde}	
\begin{split}		O^x_r&=\langle\sigma_1^x\sigma_2^y(\prod_{l=3}^{r}\sigma^z_l)\sigma_{r+1}^y\sigma_{r+2}^x\rangle\\&=\braket{B_1B_2A_{3}B_{3}\dots A_{r}B_{r}A_{r+1}A_{r+2}}.
	\end{split}	
\end{equation}
Using the same method in Ref.~\cite{WickG.C.1950}, one can rewrite $O_r^x$ as
\begin{equation}\label{ox}	O^x_r=\rm{pf}\begin{vmatrix}
		0&S_{1}&G_2&S_2&G_3&\dots&G_{r+1}\\
		 &0&G_{1}&S_1&G_{2}&\dots&G_{r}\\
		 &&0&-G_{0}&Q_1&\dots&Q_{r-1}\\
		 &&&\ddots&\ddots&&\vdots&\\
		 &&&&0&G_1&G_2\\
		 &&&&&0&Q_1\\
		 &&&&&&0\end{vmatrix}.
\end{equation}

We first study the energy gap of the ground state by using the complex energy spectra $\Delta E(k)=E_{k,+}-E_{k,-}\equiv\Delta E_{R}(k)+i\Delta E_{I}(k)$. The corresponding real and imaginary parts of the complex energy gap are given by
$\Delta_R = \min_{k\in[0,\pi]}|\Delta E_{R}(k)|$ and $\Delta_I = \min_{k\in[0,\pi]}|\Delta E_{I}(k)|$, respectively. We numerically calculate $\Delta_R$ and $\Delta_I$ in the $\lambda$-$\Gamma$ plane, as shown in Figs. \ref{fig5}(a) and \ref{fig5}(b), respectively. One can find a parameter regime for the gapless phase with $\Delta_R=\Delta_I=0$, while other regimes for gapped phases with $\Delta_R\neq0$ (real gap) or $\Delta_I\neq0$ (imaginary gap) , corresponding to line gaps in the complex energy spectra. To be more clearly, we show two typical results of $|\Delta E_{R}|$ and $|\Delta E_{I}|$ as functions of $k$ in Fig. \ref{fig5}(c), where the upper and lower sub-figures indicate the gapless and imaginary gapped phases, respectively. We then study the order parameters of the ground state. Typical numerical results of $|O_r^x|$ and $|R_r^y|$ as functions of the distance $r$ are shown in Figs. \ref{fig5}(d) and \ref{fig5}(e), respectively. Several cases for the power-law decay of $|O_r^x|$ and $|R_r^y|$ with respect to $r$ are further shown in Fig. \ref{fig5}(f).

Combing the properties of energy gaps and order parameters, one can obtain the ground-state phase diagram of the non-Hermitian cluster Ising model~\cite{GuoZheng-Xin2022}, as shown in Fig. \ref{fig5}(g). The region $\mathrm{I}$ corresponds the real gapped cluster phase with $\Delta_R\neq0$ and long-range string correlation $|O_r^x|$ and $O_x\neq0$ in the large $r$ limit. In regions $\mathrm{II}$ and $\mathrm{III}$, the ground states are respectively in the gapless and imaginary gapped cluster phases with quasi long-range string correlation, where $|O_r^x|$ exhibits the power-law decay as shown in Fig. \ref{fig5}(f). Note that the region III is overlooked in Ref.~\cite{GuoZheng-Xin2022} due to some errors in the BdG transformations and the calculation of long-range correlations. The phase in the region $\mathrm{IV}$ is the imaginary gapped antiferromagnetic phase with quasi long-range correlation of $R_r^y$ ($|R_r^y|$ exhibits the power-law decay). The region $\mathrm{V}$ corresponds to the real gapped antiferromagnetic phase with $\Delta_R\neq0$ and long-range correlation of $|R_r^y|$ and $m_y \neq 0$ in the large $r$ limit. One can further calculate the derivatives of the order parameters $O_x$ and $m_y$ with respect to the parameter $\lambda$ to determine the phase boundaries. Typical numerical results of $\partial O_x/\partial \lambda$ and $\partial m_y/\partial \lambda$ for $r=1000$ (which is large enough to approach the thermodynamic limit) for the Hermitian and non-Hermitian cases are shown in Figs. \ref{fig5}(h) and \ref{fig5}(i), respectively. The non-analytic points of the derivatives with respect to $\lambda$ correspond to the critical points. In this way, we determine the critical points from the order parameters in the whole $\lambda$-$\Gamma$ plane and plot them (blue points) in the phase diagram in Fig. \ref{fig5}(g).

To show that the critical points in the non-Hermitian cluster Ising model can be revealed by the quantum metric, we calculate $g_{\lambda\lambda}$ of the ground states as a function of $\lambda$ in Figs. \ref{fig5}(h) and \ref{fig5}(i). The locations of the peaks of $g_{\lambda\lambda}$ correspond to the critical points. We also plot these critical points (red points) in the phase diagram in Fig. \ref{fig5}(g). One can find that the critical points revealed via the quantum metric are consistent of those determined by the order parameters. However, the quantum metric is unable to witness the transition point between the quasi-long-ranged cluster (region III) and antiferromagnetic (region IV) phases, both of which have an imaginary gap. This invalidity is due to the absence of gap closing at the discontinuous transition points between phase regimes III and IV, similar to the case for the Hermitian p-wave-pair Aubry-Andr\'{e}-Harper model~\cite{LvTing2022B}. Notably, the critical points with gap closing can also be detected by the quantum metric $g_{\Gamma\Gamma}$. For instance, we plot $g_{\Gamma\Gamma}$ as a function of $\Gamma$ in Fig.~\ref{fig5}(j). One can still find the peaks at $\Gamma\approx 2.23$ with fixed $\lambda=0.5$, and at $\Gamma\approx 0.48$ and $\Gamma\approx 2.95$ with fixed $\lambda=0.9$ in $g_{\Gamma\Gamma}$, indicating one and two critical points, respectively.

\begin{figure}[tb]
	\centering
	\includegraphics[width=0.48\textwidth]{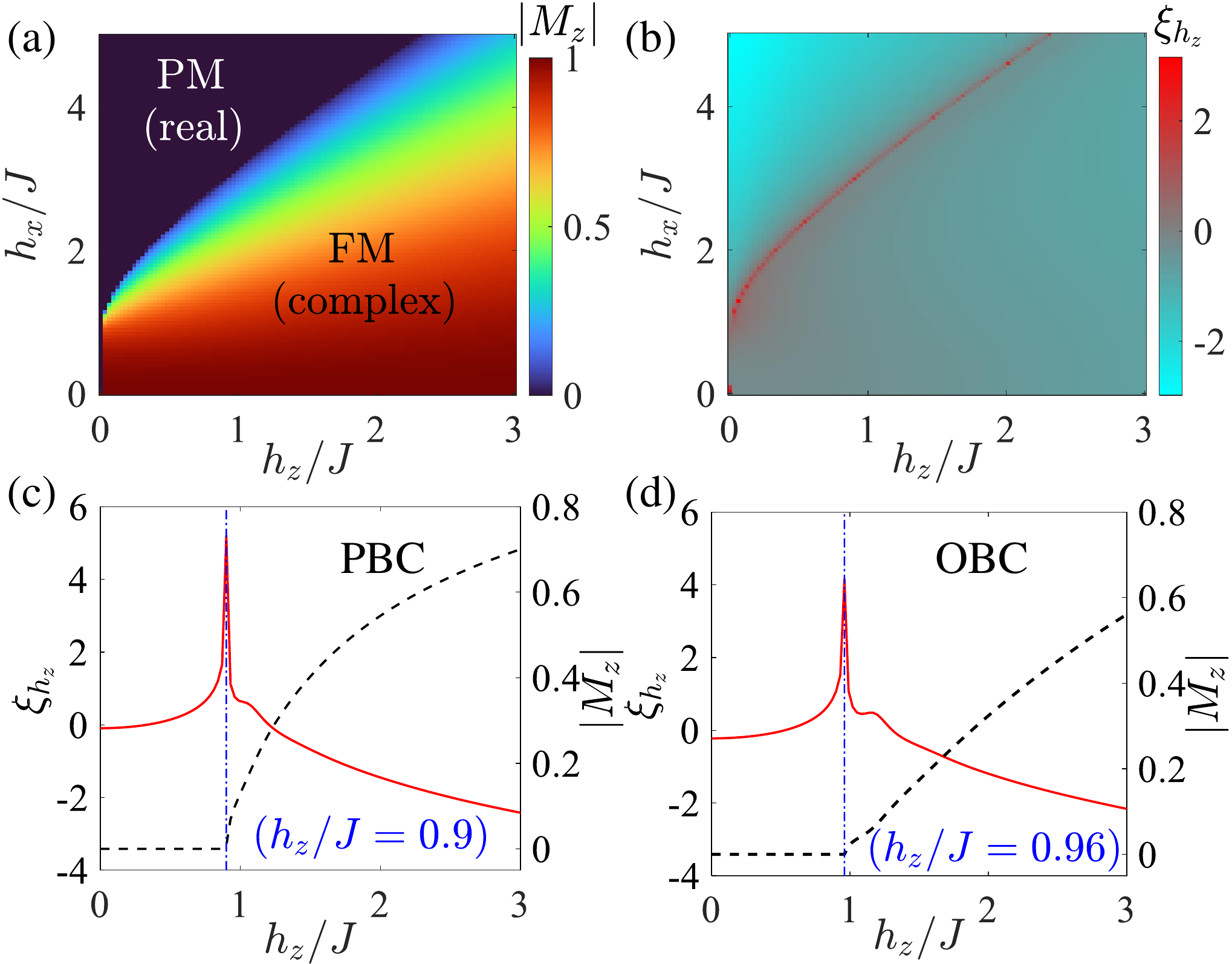}
	\caption{(Color online) The non-Hermitian mixed-field Ising model $H_{\text{NH-Ising}}^{(2)}$ with the system size $N=10$. (a) $|M_z|$ and (b) $\xi_{h_z}$ of the ground state as functions of $h_z$ and $h_x$ under the PBC. The phase diagram based on $|M_z|$ consists two regions $\mathrm{I}$ and $\mathrm{II}$ for the paramagnetic (PM) and ferromagnetic (FM) phases with real and complex energies, respectively. $|M_z|$ and $\xi_{h_z}$ as a function of $h_z$ (c) under the PBC; and (d) the OBC, respectively. The auxiliary lines at $h_z/J=0.9$ (c) and $h_z/J=0.96$ (d) are plotted, respectively. Other parameter is $J=1$.
	}\label{fig6}
\end{figure}

\subsection{Non-Hermitian mixed-field Ising model}

We further study the non-Hermitian mixed-field Ising model, which is non-integrable and described by the following Hamiltonian
\begin{equation}\label{mising}
H_{\text{NH-Ising}}^{(2)}=-J\sum_{l=1}^N \sigma_{l}^z \sigma_{l+1}^z +h_x\sum_{l=1}^N \sigma_{l}^x +ih_z\sum_{l=1}^N \sigma_{l}^z,
\end{equation}
where $h_x$ and $h_z$ denote the strengths of the real and imaginary fields that are along the transverse and longitudinal directions, respectively. The order parameter for the ground state $|\psi_1\rangle$ in this model is the magnetization $M_z=\frac{1}{N}\bra{\psi_1}\sum_{l=1}^N\sigma^z_l\ket{\psi_1}$,
where $\ket{\psi_1}$ can be numerically obtained by the exact diagonalization method for small system size $N$. The numerical results for $N=10$ under the PBC and the open boundary condition (OBC) are shown in Figs. \ref{fig6}(a-c) and Fig. \ref{fig6}(d), respectively. The phase diagram that contains the paramagnetic phase with $|M_z|\approx0$ (and real energy) and the ferromagnetic phase with finite value of $|M_z|$ (and complex energy) in the $h_z$-$h_x$ parameter plane is shown in Fig.~\ref{fig6}(a). Moreover, the phase transition with gap closing between two phases is accompanied by a real-complex transition of the ground state energy. To detect the phase transition with the quantum metric, we numerically compute the logarithm quantify of quantum metric $\xi_{h_z}=\log_{10}g_{h_zh_z}$ as a function of $h_z$ and $h_x$ in Fig~\ref{fig6}(b). One can see the boundary between the paramagnetic and ferromagnetic phases are well revealed by $\xi_{h_z}$. As shown in Fig.~\ref{fig6}(c) and Fig.~\ref{fig6}(d) for different boundary conditions with fixed $h_x=3$, the critical point $h_z\approx0.9$ and $h_z\approx0.96$ can both be clearly revealed by the quantum metric.

\section{\label{sec5} discussion and conclusion}

\begin{figure}[tb]
\centering
\includegraphics[width=0.48\textwidth]{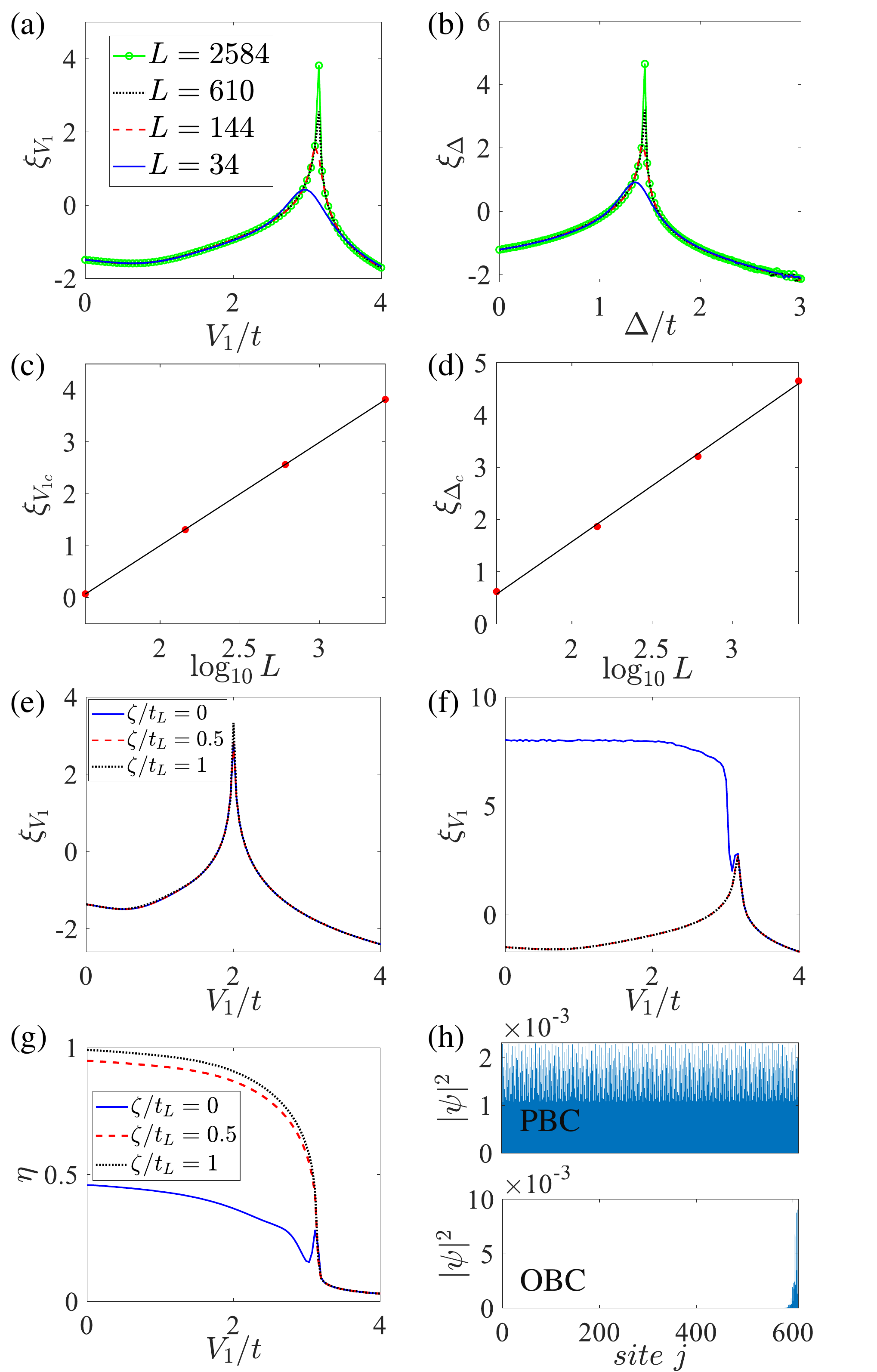}
\caption{(Color online) (a) $\xi_{V_1}$ as a function of $V_1$ for $H_{\text{GAA}}^{(1)}$ and (b) $\xi_{\Delta}$ as a function of $\Delta$ for $H_{\text{GAA}}^{(2)}$, with both $g=0.5$ and different system sizes $L$. (c) Finite-size scaling of $\xi_{V_{1c}}=\xi_{V_1}(V_1=V_c)$ obtained from (a) with $V_{1c}=3.16$ and $\xi_{V_{1c}}=1.992\log_{10}L-2.983$. (d) Finite-size scaling of $\xi_{\Delta_c}=\xi_{\Delta}(\Delta=\Delta_c)$ obtained from (b) with $\Delta_c=1.45$ and $\xi_{\Delta_c}=2.142\log_{10}L-2.708$. $\xi_{V_1}$ as a function of $V_1$ for $\tilde{H}$ [see Eq. (\ref{AAH3})] with fixed (e) $g=0$ and (f) $g=0.5$, for different boundary condition coefficients $\zeta$. (g) $\eta$ as a function of $V_1$ for different boundary condition coefficients $\zeta$ with fixed $g=0.5$. (h) The density distribution of the ground-state with fixed $V_1=0$ and $g=0.5$ under the PBC and the OBC, respectively. The system size is $L=610$. Other parameters are $h=0$, $V_2=0.5$, $\alpha =-0.5$ and $t=1$.
}\label{fig7}
\end{figure}

Before concluding, we discuss the finite-size effect and boundary conditions on detecting the non-Hermitian critical points with the quantum metric. Without loss of generality, we focus on the non-Hermitian GAA model $H_{\text{GAA}}^{(1)}$ in Eq. (\ref{AAH}) and $H_{\text{GAA}}^{(2)}$ in Eq. (\ref{GAA}). The finite-size scaling of the quantum metric of the ground state $\xi_{V_1}=\log_{10}g_{V_1V_1}$ for $H_{\text{GAA}}^{(1)}$ and $\xi_{\Delta}=\log_{10}g_{\Delta\Delta}$ for $H_{\text{GAA}}^{(2)}$ are shown in Figs.~\ref{fig7}(a) and \ref{fig7}(b), respectively. The lattice sizes $L=34,144,610,2584$ are chosen from the Fibonacci sequences to satisfy the PBC. The critical points are determined as $V_{1c}\approx3.16$ and $\Delta_c\approx1.45$, respectively. We also plot the quantum metric at the critical points $\xi_{V_{1c}}=\xi_{V_1}(V_1=V_c)$ and $\xi_{\Delta_c}=\xi_{\Delta}(\Delta=\Delta_c)$ versus $L$ in Figs.~\ref{fig7}(c) and \ref{fig7}(d), respectively. One can find that the quantum metric at the critical points exhibits the scaling law $g_{\mu\mu}\sim L^{\kappa}$ (i.e., $\xi_{\mu\mu}\sim\kappa \log_{10}L$ up to a constant), where the exponent $\kappa$ is related to the critical exponents~\cite{GU2010,LvTing2022A,LvTing2022B}. The critical exponents for the non-Hermitian localization and quantum phase transitions may be extracted from the scaling analysis of the quantum metric and order parameters, which require further investigations.

To consider the boundary effect on the quantum metric, we add an boundary coupling term to the Hamiltonian $H_{\text{GAA}}^{(1)}$ as
\begin{equation}\label{AAH3}
	\begin{split}
		\tilde{H}=H_{\text{GAA}}^{(1)}+\zeta(e^{-g} c^\dagger_{1} c_L+e^{g} c^\dagger_L c_{1}),
	\end{split}
\end{equation}
where $\zeta$ is the coefficient for tuning different boundary conditions. For $\zeta=0$, it denotes the open boundary condition, while $\zeta=1$ ($\zeta\neq0,1$) denotes the standard (modified) periodic boundary condition. As shown in Fig.~\ref{fig7}(e), the results of $\xi_{V_1}$ as a function of $V_1$ under different boundary conditions ($\zeta=0,0.5,1$) coincide in the Hermitian limit. In the non-Hermitian case, the results of $\xi_{V_1}$ are different for periodic ($\zeta=0.5,1$) and open ($\zeta=0$) boundary conditions, while the transition point can revealed in both conditions, as shown in Fig.~\ref{fig7}(f). The difference comes from the localized bulk eigenstates induced by the non-Hermitian skin effect under the open boundary condition \cite{Yao2018,Songfei2019}, as shown in Figs.~\ref{fig7}(g) and (h). The non-Hermitian skin effect destroys the extended phase under the original periodic boundary condition and leads to the increase of the quantum metric with accidental degeneracies. Thus, our results are robust against to the finite-size effect and different boundary conditions when using the quantum metric to reveal critical points.

In summary, we have investigated the quantum phase transitions in various non-Hermitian systems by using the information-geometric approach. It has been shown that the peaks of the quantum metric of the eigenstates exactly identify the localization transition points and mobility edges in the non-Hermitian GAA models. For non-Hermitian cluster and mixed-field Ising models, we have demonstrated that the phase boundaries in the non-Hermiticity parameter space determined by quantum metric of ground state perfectly coincide with those from the corresponding order parameters. These results indicate that the peak of the quantum metric serving as good signatures for detecting the non-Hermitian critical points with gap closings in both single-particle quasi-periodic and many-body spin systems. This strategy is robust against the finite-size effect and different boundary conditions. Notably, our present work only focuses on the right eigenstates of non-Hermitian Hamiltonians, and the quantum metric effect based on the biorthogonal eigenstates and the non-unitary dynamics are deserving to be further explored.

-\textit{Note added.} After finishing our manuscript, we noticed two relevant preprints~\cite{orlov2024adiabatic,zeng2024fidelity}. In Ref.~\cite{orlov2024adiabatic}, a generalization of the QGT based on the generator of adiabatic transformations was developed to study phase transitions in non-Hermitian quantum systems. In Ref.~\cite{zeng2024fidelity}, the self-normal and biorthogonal fidelity susceptibilities were used to study the critical behaviors in the nonreciprocal Aubry-Andr\'{e}-Harper model.

\begin{acknowledgments}
We thank Zhi Li for helpful discussions. This work was supported by the National Natural Science Foundation of China (Grant No. 12174126), the Guangdong Basic and Applied Basic Research Foundation (Grant No. 2024B1515020018), and the Science and Technology Program of Guangzhou (Grant No. 2024A04J3004).
\end{acknowledgments}
	
	\bibliography{reference}

\begin{thebibliography}{113}%
\makeatletter
\providecommand \@ifxundefined [1]{%
 \@ifx{#1\undefined}
}%
\providecommand \@ifnum [1]{%
 \ifnum #1\expandafter \@firstoftwo
 \else \expandafter \@secondoftwo
 \fi
}%
\providecommand \@ifx [1]{%
 \ifx #1\expandafter \@firstoftwo
 \else \expandafter \@secondoftwo
 \fi
}%
\providecommand \natexlab [1]{#1}%
\providecommand \enquote  [1]{``#1''}%
\providecommand \bibnamefont  [1]{#1}%
\providecommand \bibfnamefont [1]{#1}%
\providecommand \citenamefont [1]{#1}%
\providecommand \href@noop [0]{\@secondoftwo}%
\providecommand \href [0]{\begingroup \@sanitize@url \@href}%
\providecommand \@href[1]{\@@startlink{#1}\@@href}%
\providecommand \@@href[1]{\endgroup#1\@@endlink}%
\providecommand \@sanitize@url [0]{\catcode `\\12\catcode `\$12\catcode
  `\&12\catcode `\#12\catcode `\^12\catcode `\_12\catcode `\%12\relax}%
\providecommand \@@startlink[1]{}%
\providecommand \@@endlink[0]{}%
\providecommand \url  [0]{\begingroup\@sanitize@url \@url }%
\providecommand \@url [1]{\endgroup\@href {#1}{\urlprefix }}%
\providecommand \urlprefix  [0]{URL }%
\providecommand \Eprint [0]{\href }%
\providecommand \doibase [0]{http://dx.doi.org/}%
\providecommand \selectlanguage [0]{\@gobble}%
\providecommand \bibinfo  [0]{\@secondoftwo}%
\providecommand \bibfield  [0]{\@secondoftwo}%
\providecommand \translation [1]{[#1]}%
\providecommand \BibitemOpen [0]{}%
\providecommand \bibitemStop [0]{}%
\providecommand \bibitemNoStop [0]{.\EOS\space}%
\providecommand \EOS [0]{\spacefactor3000\relax}%
\providecommand \BibitemShut  [1]{\csname bibitem#1\endcsname}%
\let\auto@bib@innerbib\@empty
\bibitem [{\citenamefont {Sachdev}(2011)}]{Sachdev_2011}%
  \BibitemOpen
  \bibfield  {author} {\bibinfo {author} {\bibfnamefont {S.}~\bibnamefont
  {Sachdev}},\ }\href@noop {} {\emph {\bibinfo {title} {Quantum Phase
  Transitions}}},\ \bibinfo {edition} {2nd}\ ed.\ (\bibinfo  {publisher}
  {Cambridge University Press},\ \bibinfo {year} {2011})\BibitemShut {NoStop}%
\bibitem [{\citenamefont {Smacchia}\ \emph {et~al.}(2011)\citenamefont
  {Smacchia}, \citenamefont {Amico}, \citenamefont {Facchi}, \citenamefont
  {Fazio}, \citenamefont {Florio}, \citenamefont {Pascazio},\ and\
  \citenamefont {Vedral}}]{Smacchia2011}%
  \BibitemOpen
  \bibfield  {author} {\bibinfo {author} {\bibfnamefont {P.}~\bibnamefont
  {Smacchia}}, \bibinfo {author} {\bibfnamefont {L.}~\bibnamefont {Amico}},
  \bibinfo {author} {\bibfnamefont {P.}~\bibnamefont {Facchi}}, \bibinfo
  {author} {\bibfnamefont {R.}~\bibnamefont {Fazio}}, \bibinfo {author}
  {\bibfnamefont {G.}~\bibnamefont {Florio}}, \bibinfo {author} {\bibfnamefont
  {S.}~\bibnamefont {Pascazio}}, \ and\ \bibinfo {author} {\bibfnamefont
  {V.}~\bibnamefont {Vedral}},\ }\href {\doibase 10.1103/PhysRevA.84.022304}
  {\bibfield  {journal} {\bibinfo  {journal} {Phys. Rev. A}\ }\textbf {\bibinfo
  {volume} {84}},\ \bibinfo {pages} {022304} (\bibinfo {year}
  {2011})}\BibitemShut {NoStop}%
\bibitem [{\citenamefont {Ganeshan}\ \emph {et~al.}(2015)\citenamefont
  {Ganeshan}, \citenamefont {Pixley},\ and\ \citenamefont
  {Das~Sarma}}]{Ganeshan2015}%
  \BibitemOpen
  \bibfield  {author} {\bibinfo {author} {\bibfnamefont {S.}~\bibnamefont
  {Ganeshan}}, \bibinfo {author} {\bibfnamefont {J.~H.}\ \bibnamefont
  {Pixley}}, \ and\ \bibinfo {author} {\bibfnamefont {S.}~\bibnamefont
  {Das~Sarma}},\ }\href {\doibase 10.1103/PhysRevLett.114.146601} {\bibfield
  {journal} {\bibinfo  {journal} {Phys. Rev. Lett.}\ }\textbf {\bibinfo
  {volume} {114}},\ \bibinfo {pages} {146601} (\bibinfo {year}
  {2015})}\BibitemShut {NoStop}%
\bibitem [{\citenamefont {An}\ \emph {et~al.}(2021)\citenamefont {An},
  \citenamefont {Padavi\ifmmode~\acute{c}\else \'{c}\fi{}}, \citenamefont
  {Meier}, \citenamefont {Hegde}, \citenamefont {Ganeshan}, \citenamefont
  {Pixley}, \citenamefont {Vishveshwara},\ and\ \citenamefont
  {Gadway}}]{An2021}%
  \BibitemOpen
  \bibfield  {author} {\bibinfo {author} {\bibfnamefont {F.~A.}\ \bibnamefont
  {An}}, \bibinfo {author} {\bibfnamefont {K.}~\bibnamefont
  {Padavi\ifmmode~\acute{c}\else \'{c}\fi{}}}, \bibinfo {author} {\bibfnamefont
  {E.~J.}\ \bibnamefont {Meier}}, \bibinfo {author} {\bibfnamefont
  {S.}~\bibnamefont {Hegde}}, \bibinfo {author} {\bibfnamefont
  {S.}~\bibnamefont {Ganeshan}}, \bibinfo {author} {\bibfnamefont {J.~H.}\
  \bibnamefont {Pixley}}, \bibinfo {author} {\bibfnamefont {S.}~\bibnamefont
  {Vishveshwara}}, \ and\ \bibinfo {author} {\bibfnamefont {B.}~\bibnamefont
  {Gadway}},\ }\href {\doibase 10.1103/PhysRevLett.126.040603} {\bibfield
  {journal} {\bibinfo  {journal} {Phys. Rev. Lett.}\ }\textbf {\bibinfo
  {volume} {126}},\ \bibinfo {pages} {040603} (\bibinfo {year}
  {2021})}\BibitemShut {NoStop}%
\bibitem [{\citenamefont {Wang}\ \emph {et~al.}(2022)\citenamefont {Wang},
  \citenamefont {Zhang}, \citenamefont {Li}, \citenamefont {Wu}, \citenamefont
  {Liu}, \citenamefont {Mei}, \citenamefont {Hu}, \citenamefont {Xiao},
  \citenamefont {Ma}, \citenamefont {Chin},\ and\ \citenamefont
  {Jia}}]{WangYunfei2022}%
  \BibitemOpen
  \bibfield  {author} {\bibinfo {author} {\bibfnamefont {Y.}~\bibnamefont
  {Wang}}, \bibinfo {author} {\bibfnamefont {J.-H.}\ \bibnamefont {Zhang}},
  \bibinfo {author} {\bibfnamefont {Y.}~\bibnamefont {Li}}, \bibinfo {author}
  {\bibfnamefont {J.}~\bibnamefont {Wu}}, \bibinfo {author} {\bibfnamefont
  {W.}~\bibnamefont {Liu}}, \bibinfo {author} {\bibfnamefont {F.}~\bibnamefont
  {Mei}}, \bibinfo {author} {\bibfnamefont {Y.}~\bibnamefont {Hu}}, \bibinfo
  {author} {\bibfnamefont {L.}~\bibnamefont {Xiao}}, \bibinfo {author}
  {\bibfnamefont {J.}~\bibnamefont {Ma}}, \bibinfo {author} {\bibfnamefont
  {C.}~\bibnamefont {Chin}}, \ and\ \bibinfo {author} {\bibfnamefont
  {S.}~\bibnamefont {Jia}},\ }\href {\doibase 10.1103/PhysRevLett.129.103401}
  {\bibfield  {journal} {\bibinfo  {journal} {Phys. Rev. Lett.}\ }\textbf
  {\bibinfo {volume} {129}},\ \bibinfo {pages} {103401} (\bibinfo {year}
  {2022})}\BibitemShut {NoStop}%
\bibitem [{\citenamefont {Tang}\ \emph {et~al.}(2021)\citenamefont {Tang},
  \citenamefont {Zhang}, \citenamefont {Zhang},\ and\ \citenamefont
  {Zhang}}]{TangLingZhi2021}%
  \BibitemOpen
  \bibfield  {author} {\bibinfo {author} {\bibfnamefont {L.-Z.}\ \bibnamefont
  {Tang}}, \bibinfo {author} {\bibfnamefont {G.-Q.}\ \bibnamefont {Zhang}},
  \bibinfo {author} {\bibfnamefont {L.-F.}\ \bibnamefont {Zhang}}, \ and\
  \bibinfo {author} {\bibfnamefont {D.-W.}\ \bibnamefont {Zhang}},\ }\href
  {\doibase 10.1103/PhysRevA.103.033325} {\bibfield  {journal} {\bibinfo
  {journal} {Phys. Rev. A}\ }\textbf {\bibinfo {volume} {103}},\ \bibinfo
  {pages} {033325} (\bibinfo {year} {2021})}\BibitemShut {NoStop}%
\bibitem [{\citenamefont {Matsumoto}\ \emph {et~al.}(2020)\citenamefont
  {Matsumoto}, \citenamefont {Kawabata}, \citenamefont {Ashida}, \citenamefont
  {Furukawa},\ and\ \citenamefont {Ueda}}]{Matsumoto2020}%
  \BibitemOpen
  \bibfield  {author} {\bibinfo {author} {\bibfnamefont {N.}~\bibnamefont
  {Matsumoto}}, \bibinfo {author} {\bibfnamefont {K.}~\bibnamefont {Kawabata}},
  \bibinfo {author} {\bibfnamefont {Y.}~\bibnamefont {Ashida}}, \bibinfo
  {author} {\bibfnamefont {S.}~\bibnamefont {Furukawa}}, \ and\ \bibinfo
  {author} {\bibfnamefont {M.}~\bibnamefont {Ueda}},\ }\href {\doibase
  10.1103/PhysRevLett.125.260601} {\bibfield  {journal} {\bibinfo  {journal}
  {Phys. Rev. Lett.}\ }\textbf {\bibinfo {volume} {125}},\ \bibinfo {pages}
  {260601} (\bibinfo {year} {2020})}\BibitemShut {NoStop}%
\bibitem [{\citenamefont {Longhi}(2019)}]{Longhi2019}%
  \BibitemOpen
  \bibfield  {author} {\bibinfo {author} {\bibfnamefont {S.}~\bibnamefont
  {Longhi}},\ }\href {\doibase 10.1103/PhysRevLett.122.237601} {\bibfield
  {journal} {\bibinfo  {journal} {Phys. Rev. Lett.}\ }\textbf {\bibinfo
  {volume} {122}},\ \bibinfo {pages} {237601} (\bibinfo {year}
  {2019})}\BibitemShut {NoStop}%
\bibitem [{\citenamefont {Gong}\ \emph {et~al.}(2018)\citenamefont {Gong},
  \citenamefont {Ashida}, \citenamefont {Kawabata}, \citenamefont {Takasan},
  \citenamefont {Higashikawa},\ and\ \citenamefont {Ueda}}]{GongZongping2018}%
  \BibitemOpen
  \bibfield  {author} {\bibinfo {author} {\bibfnamefont {Z.}~\bibnamefont
  {Gong}}, \bibinfo {author} {\bibfnamefont {Y.}~\bibnamefont {Ashida}},
  \bibinfo {author} {\bibfnamefont {K.}~\bibnamefont {Kawabata}}, \bibinfo
  {author} {\bibfnamefont {K.}~\bibnamefont {Takasan}}, \bibinfo {author}
  {\bibfnamefont {S.}~\bibnamefont {Higashikawa}}, \ and\ \bibinfo {author}
  {\bibfnamefont {M.}~\bibnamefont {Ueda}},\ }\href {\doibase
  10.1103/PhysRevX.8.031079} {\bibfield  {journal} {\bibinfo  {journal} {Phys.
  Rev. X}\ }\textbf {\bibinfo {volume} {8}},\ \bibinfo {pages} {031079}
  (\bibinfo {year} {2018})}\BibitemShut {NoStop}%
\bibitem [{\citenamefont {Zhang}\ \emph {et~al.}(2021)\citenamefont {Zhang},
  \citenamefont {Tang}, \citenamefont {Zhang}, \citenamefont {Zhang},\ and\
  \citenamefont {Zhu}}]{ZhangGuo-Qing2021}%
  \BibitemOpen
  \bibfield  {author} {\bibinfo {author} {\bibfnamefont {G.-Q.}\ \bibnamefont
  {Zhang}}, \bibinfo {author} {\bibfnamefont {L.-Z.}\ \bibnamefont {Tang}},
  \bibinfo {author} {\bibfnamefont {L.-F.}\ \bibnamefont {Zhang}}, \bibinfo
  {author} {\bibfnamefont {D.-W.}\ \bibnamefont {Zhang}}, \ and\ \bibinfo
  {author} {\bibfnamefont {S.-L.}\ \bibnamefont {Zhu}},\ }\href {\doibase
  10.1103/PhysRevB.104.L161118} {\bibfield  {journal} {\bibinfo  {journal}
  {Phys. Rev. B}\ }\textbf {\bibinfo {volume} {104}},\ \bibinfo {pages}
  {L161118} (\bibinfo {year} {2021})}\BibitemShut {NoStop}%
\bibitem [{\citenamefont {Zanardi}\ and\ \citenamefont
  {Paunkovi\ifmmode~\acute{c}\else \'{c}\fi{}}(2006)}]{ZanardiPaolo2006}%
  \BibitemOpen
  \bibfield  {author} {\bibinfo {author} {\bibfnamefont {P.}~\bibnamefont
  {Zanardi}}\ and\ \bibinfo {author} {\bibfnamefont {N.}~\bibnamefont
  {Paunkovi\ifmmode~\acute{c}\else \'{c}\fi{}}},\ }\href {\doibase
  10.1103/PhysRevE.74.031123} {\bibfield  {journal} {\bibinfo  {journal} {Phys.
  Rev. E}\ }\textbf {\bibinfo {volume} {74}},\ \bibinfo {pages} {031123}
  (\bibinfo {year} {2006})}\BibitemShut {NoStop}%
\bibitem [{\citenamefont {Zhu}(2006)}]{Zhu2006}%
  \BibitemOpen
  \bibfield  {author} {\bibinfo {author} {\bibfnamefont {S.-L.}\ \bibnamefont
  {Zhu}},\ }\href {\doibase 10.1103/PhysRevLett.96.077206} {\bibfield
  {journal} {\bibinfo  {journal} {Phys. Rev. Lett.}\ }\textbf {\bibinfo
  {volume} {96}},\ \bibinfo {pages} {077206} (\bibinfo {year}
  {2006})}\BibitemShut {NoStop}%
\bibitem [{\citenamefont {Zhu}(2008)}]{Zhu2008}%
  \BibitemOpen
  \bibfield  {author} {\bibinfo {author} {\bibfnamefont {S.-L.}\ \bibnamefont
  {Zhu}},\ }\href {\doibase 10.1142/S0217979208038855} {\bibfield  {journal}
  {\bibinfo  {journal} {Int. J. Mod. Phys. B}\ }\textbf {\bibinfo {volume}
  {22}},\ \bibinfo {pages} {561} (\bibinfo {year} {2008})}\BibitemShut
  {NoStop}%
\bibitem [{\citenamefont {Zanardi}\ \emph
  {et~al.}(2007{\natexlab{a}})\citenamefont {Zanardi}, \citenamefont {Giorda},\
  and\ \citenamefont {Cozzini}}]{zanardi2007information}%
  \BibitemOpen
  \bibfield  {author} {\bibinfo {author} {\bibfnamefont {P.}~\bibnamefont
  {Zanardi}}, \bibinfo {author} {\bibfnamefont {P.}~\bibnamefont {Giorda}}, \
  and\ \bibinfo {author} {\bibfnamefont {M.}~\bibnamefont {Cozzini}},\ }\href
  {\doibase 10.1103/PhysRevLett.99.100603} {\bibfield  {journal} {\bibinfo
  {journal} {Phys. Rev. Lett.}\ }\textbf {\bibinfo {volume} {99}},\ \bibinfo
  {pages} {100603} (\bibinfo {year} {2007}{\natexlab{a}})}\BibitemShut
  {NoStop}%
\bibitem [{\citenamefont {You}\ \emph {et~al.}(2007)\citenamefont {You},
  \citenamefont {Li},\ and\ \citenamefont {Gu}}]{YouWen-Long2007}%
  \BibitemOpen
  \bibfield  {author} {\bibinfo {author} {\bibfnamefont {W.-L.}\ \bibnamefont
  {You}}, \bibinfo {author} {\bibfnamefont {Y.-W.}\ \bibnamefont {Li}}, \ and\
  \bibinfo {author} {\bibfnamefont {S.-J.}\ \bibnamefont {Gu}},\ }\href
  {\doibase 10.1103/PhysRevE.76.022101} {\bibfield  {journal} {\bibinfo
  {journal} {Phys. Rev. E}\ }\textbf {\bibinfo {volume} {76}},\ \bibinfo
  {pages} {022101} (\bibinfo {year} {2007})}\BibitemShut {NoStop}%
\bibitem [{\citenamefont {Gu}(2010)}]{GU2010}%
  \BibitemOpen
  \bibfield  {author} {\bibinfo {author} {\bibfnamefont {S.-J.}\ \bibnamefont
  {Gu}},\ }\href {\doibase 10.1142/S0217979210056335} {\bibfield  {journal}
  {\bibinfo  {journal} {Int. J. Mod. Phys. B}\ }\textbf {\bibinfo {volume}
  {24}},\ \bibinfo {pages} {4371} (\bibinfo {year} {2010})}\BibitemShut
  {NoStop}%
\bibitem [{\citenamefont {Lv}\ \emph {et~al.}(2022{\natexlab{a}})\citenamefont
  {Lv}, \citenamefont {Yi}, \citenamefont {Li}, \citenamefont {Sun},\ and\
  \citenamefont {You}}]{LvTing2022A}%
  \BibitemOpen
  \bibfield  {author} {\bibinfo {author} {\bibfnamefont {T.}~\bibnamefont
  {Lv}}, \bibinfo {author} {\bibfnamefont {T.-C.}\ \bibnamefont {Yi}}, \bibinfo
  {author} {\bibfnamefont {L.}~\bibnamefont {Li}}, \bibinfo {author}
  {\bibfnamefont {G.}~\bibnamefont {Sun}}, \ and\ \bibinfo {author}
  {\bibfnamefont {W.-L.}\ \bibnamefont {You}},\ }\href {\doibase
  10.1103/PhysRevA.105.013315} {\bibfield  {journal} {\bibinfo  {journal}
  {Phys. Rev. A}\ }\textbf {\bibinfo {volume} {105}},\ \bibinfo {pages}
  {013315} (\bibinfo {year} {2022}{\natexlab{a}})}\BibitemShut {NoStop}%
\bibitem [{\citenamefont {Lv}\ \emph {et~al.}(2022{\natexlab{b}})\citenamefont
  {Lv}, \citenamefont {Liu}, \citenamefont {Yi}, \citenamefont {Li},
  \citenamefont {Liu},\ and\ \citenamefont {You}}]{LvTing2022B}%
  \BibitemOpen
  \bibfield  {author} {\bibinfo {author} {\bibfnamefont {T.}~\bibnamefont
  {Lv}}, \bibinfo {author} {\bibfnamefont {Y.-B.}\ \bibnamefont {Liu}},
  \bibinfo {author} {\bibfnamefont {T.-C.}\ \bibnamefont {Yi}}, \bibinfo
  {author} {\bibfnamefont {L.}~\bibnamefont {Li}}, \bibinfo {author}
  {\bibfnamefont {M.}~\bibnamefont {Liu}}, \ and\ \bibinfo {author}
  {\bibfnamefont {W.-L.}\ \bibnamefont {You}},\ }\href {\doibase
  10.1103/PhysRevB.106.144205} {\bibfield  {journal} {\bibinfo  {journal}
  {Phys. Rev. B}\ }\textbf {\bibinfo {volume} {106}},\ \bibinfo {pages}
  {144205} (\bibinfo {year} {2022}{\natexlab{b}})}\BibitemShut {NoStop}%
\bibitem [{\citenamefont {Ma}\ \emph {et~al.}(2010)\citenamefont {Ma},
  \citenamefont {Chen}, \citenamefont {Fan},\ and\ \citenamefont
  {Liu}}]{ma2010abelian}%
  \BibitemOpen
  \bibfield  {author} {\bibinfo {author} {\bibfnamefont {Y.-Q.}\ \bibnamefont
  {Ma}}, \bibinfo {author} {\bibfnamefont {S.}~\bibnamefont {Chen}}, \bibinfo
  {author} {\bibfnamefont {H.}~\bibnamefont {Fan}}, \ and\ \bibinfo {author}
  {\bibfnamefont {W.-M.}\ \bibnamefont {Liu}},\ }\href {\doibase
  10.1103/PhysRevB.81.245129} {\bibfield  {journal} {\bibinfo  {journal} {Phys.
  Rev. B}\ }\textbf {\bibinfo {volume} {81}},\ \bibinfo {pages} {245129}
  (\bibinfo {year} {2010})}\BibitemShut {NoStop}%
\bibitem [{\citenamefont {Carollo}\ \emph {et~al.}(2020)\citenamefont
  {Carollo}, \citenamefont {Valenti},\ and\ \citenamefont
  {Spagnolo}}]{Carollo2020}%
  \BibitemOpen
  \bibfield  {author} {\bibinfo {author} {\bibfnamefont {A.}~\bibnamefont
  {Carollo}}, \bibinfo {author} {\bibfnamefont {D.}~\bibnamefont {Valenti}}, \
  and\ \bibinfo {author} {\bibfnamefont {B.}~\bibnamefont {Spagnolo}},\ }\href
  {\doibase https://doi.org/10.1016/j.physrep.2019.11.002} {\bibfield
  {journal} {\bibinfo  {journal} {Phys. Rep.}\ }\textbf {\bibinfo {volume}
  {838}},\ \bibinfo {pages} {1} (\bibinfo {year} {2020})}\BibitemShut {NoStop}%
\bibitem [{\citenamefont {Kolodrubetz}\ \emph {et~al.}(2017)\citenamefont
  {Kolodrubetz}, \citenamefont {Sels}, \citenamefont {Mehta},\ and\
  \citenamefont {Polkovnikov}}]{Michael2017}%
  \BibitemOpen
  \bibfield  {author} {\bibinfo {author} {\bibfnamefont {M.}~\bibnamefont
  {Kolodrubetz}}, \bibinfo {author} {\bibfnamefont {D.}~\bibnamefont {Sels}},
  \bibinfo {author} {\bibfnamefont {P.}~\bibnamefont {Mehta}}, \ and\ \bibinfo
  {author} {\bibfnamefont {A.}~\bibnamefont {Polkovnikov}},\ }\href {\doibase
  https://doi.org/10.1016/j.physrep.2017.07.001} {\bibfield  {journal}
  {\bibinfo  {journal} {Phys. Rep.}\ }\textbf {\bibinfo {volume} {697}},\
  \bibinfo {pages} {1} (\bibinfo {year} {2017})}\BibitemShut {NoStop}%
\bibitem [{\citenamefont {Ding}\ \emph {et~al.}(2024)\citenamefont {Ding},
  \citenamefont {Zhang}, \citenamefont {Liu}, \citenamefont {Wang},
  \citenamefont {Zhang},\ and\ \citenamefont {Zhu}}]{ding2024non}%
  \BibitemOpen
  \bibfield  {author} {\bibinfo {author} {\bibfnamefont {H.-T.}\ \bibnamefont
  {Ding}}, \bibinfo {author} {\bibfnamefont {C.-X.}\ \bibnamefont {Zhang}},
  \bibinfo {author} {\bibfnamefont {J.-X.}\ \bibnamefont {Liu}}, \bibinfo
  {author} {\bibfnamefont {J.-T.}\ \bibnamefont {Wang}}, \bibinfo {author}
  {\bibfnamefont {D.-W.}\ \bibnamefont {Zhang}}, \ and\ \bibinfo {author}
  {\bibfnamefont {S.-L.}\ \bibnamefont {Zhu}},\ }\href {\doibase
  10.1103/PhysRevA.109.043305} {\bibfield  {journal} {\bibinfo  {journal}
  {Phys. Rev. A}\ }\textbf {\bibinfo {volume} {109}},\ \bibinfo {pages}
  {043305} (\bibinfo {year} {2024})}\BibitemShut {NoStop}%
\bibitem [{\citenamefont {Ding}\ \emph {et~al.}(2022)\citenamefont {Ding},
  \citenamefont {Zhu}, \citenamefont {He}, \citenamefont {Liu}, \citenamefont
  {Wang}, \citenamefont {Zhang},\ and\ \citenamefont {Zhu}}]{DingHai-Tao2022}%
  \BibitemOpen
  \bibfield  {author} {\bibinfo {author} {\bibfnamefont {H.-T.}\ \bibnamefont
  {Ding}}, \bibinfo {author} {\bibfnamefont {Y.-Q.}\ \bibnamefont {Zhu}},
  \bibinfo {author} {\bibfnamefont {P.}~\bibnamefont {He}}, \bibinfo {author}
  {\bibfnamefont {Y.-G.}\ \bibnamefont {Liu}}, \bibinfo {author} {\bibfnamefont
  {J.-T.}\ \bibnamefont {Wang}}, \bibinfo {author} {\bibfnamefont {D.-W.}\
  \bibnamefont {Zhang}}, \ and\ \bibinfo {author} {\bibfnamefont {S.-L.}\
  \bibnamefont {Zhu}},\ }\href {\doibase 10.1103/PhysRevA.105.012210}
  {\bibfield  {journal} {\bibinfo  {journal} {Phys. Rev. A}\ }\textbf {\bibinfo
  {volume} {105}},\ \bibinfo {pages} {012210} (\bibinfo {year}
  {2022})}\BibitemShut {NoStop}%
\bibitem [{\citenamefont {T\"orm\"a}(2023)}]{Essay2023}%
  \BibitemOpen
  \bibfield  {author} {\bibinfo {author} {\bibfnamefont {P.}~\bibnamefont
  {T\"orm\"a}},\ }\href {\doibase 10.1103/PhysRevLett.131.240001} {\bibfield
  {journal} {\bibinfo  {journal} {Phys. Rev. Lett.}\ }\textbf {\bibinfo
  {volume} {131}},\ \bibinfo {pages} {240001} (\bibinfo {year}
  {2023})}\BibitemShut {NoStop}%
\bibitem [{\citenamefont {Provost}\ and\ \citenamefont
  {Vallee}(1980)}]{provost1980}%
  \BibitemOpen
  \bibfield  {author} {\bibinfo {author} {\bibfnamefont {J.}~\bibnamefont
  {Provost}}\ and\ \bibinfo {author} {\bibfnamefont {G.}~\bibnamefont
  {Vallee}},\ }\href {\doibase 10.1007/BF02193559} {\bibfield  {journal}
  {\bibinfo  {journal} {Commun. Math. Phys.}\ }\textbf {\bibinfo {volume}
  {76}},\ \bibinfo {pages} {289} (\bibinfo {year} {1980})}\BibitemShut
  {NoStop}%
\bibitem [{\citenamefont {Thouless}\ \emph {et~al.}(1982)\citenamefont
  {Thouless}, \citenamefont {Kohmoto}, \citenamefont {Nightingale},\ and\
  \citenamefont {den Nijs}}]{Thouless1982}%
  \BibitemOpen
  \bibfield  {author} {\bibinfo {author} {\bibfnamefont {D.~J.}\ \bibnamefont
  {Thouless}}, \bibinfo {author} {\bibfnamefont {M.}~\bibnamefont {Kohmoto}},
  \bibinfo {author} {\bibfnamefont {M.~P.}\ \bibnamefont {Nightingale}}, \ and\
  \bibinfo {author} {\bibfnamefont {M.}~\bibnamefont {den Nijs}},\ }\href
  {\doibase 10.1103/PhysRevLett.49.405} {\bibfield  {journal} {\bibinfo
  {journal} {Phys. Rev. Lett.}\ }\textbf {\bibinfo {volume} {49}},\ \bibinfo
  {pages} {405} (\bibinfo {year} {1982})}\BibitemShut {NoStop}%
\bibitem [{\citenamefont {Simon}(1983)}]{Simon1983}%
  \BibitemOpen
  \bibfield  {author} {\bibinfo {author} {\bibfnamefont {B.}~\bibnamefont
  {Simon}},\ }\href {\doibase 10.1103/PhysRevLett.51.2167} {\bibfield
  {journal} {\bibinfo  {journal} {Phys. Rev. Lett.}\ }\textbf {\bibinfo
  {volume} {51}},\ \bibinfo {pages} {2167} (\bibinfo {year}
  {1983})}\BibitemShut {NoStop}%
\bibitem [{\citenamefont {Berry}(1984)}]{berry1984}%
  \BibitemOpen
  \bibfield  {author} {\bibinfo {author} {\bibfnamefont {M.}~\bibnamefont
  {Berry}},\ }\href {\doibase 10.1098/rspa.1984.0023} {\bibfield  {journal}
  {\bibinfo  {journal} {Proc. R. Soc. Lond. A}\ }\textbf {\bibinfo {volume}
  {392}},\ \bibinfo {pages} {45} (\bibinfo {year} {1984})}\BibitemShut
  {NoStop}%
\bibitem [{\citenamefont {Guo}\ \emph {et~al.}(2016)\citenamefont {Guo},
  \citenamefont {Zhong}, \citenamefont {Jing}, \citenamefont {Fu},\ and\
  \citenamefont {Wang}}]{GuoWei2016}%
  \BibitemOpen
  \bibfield  {author} {\bibinfo {author} {\bibfnamefont {W.}~\bibnamefont
  {Guo}}, \bibinfo {author} {\bibfnamefont {W.}~\bibnamefont {Zhong}}, \bibinfo
  {author} {\bibfnamefont {X.-X.}\ \bibnamefont {Jing}}, \bibinfo {author}
  {\bibfnamefont {L.-B.}\ \bibnamefont {Fu}}, \ and\ \bibinfo {author}
  {\bibfnamefont {X.}~\bibnamefont {Wang}},\ }\href {\doibase
  10.1103/PhysRevA.93.042115} {\bibfield  {journal} {\bibinfo  {journal} {Phys.
  Rev. A}\ }\textbf {\bibinfo {volume} {93}},\ \bibinfo {pages} {042115}
  (\bibinfo {year} {2016})}\BibitemShut {NoStop}%
\bibitem [{\citenamefont {Sundaram}\ and\ \citenamefont
  {Niu}(1999)}]{sundaram1999wave}%
  \BibitemOpen
  \bibfield  {author} {\bibinfo {author} {\bibfnamefont {G.}~\bibnamefont
  {Sundaram}}\ and\ \bibinfo {author} {\bibfnamefont {Q.}~\bibnamefont {Niu}},\
  }\href {\doibase 10.1103/PhysRevB.59.14915} {\bibfield  {journal} {\bibinfo
  {journal} {Phys. Rev. B}\ }\textbf {\bibinfo {volume} {59}},\ \bibinfo
  {pages} {14915} (\bibinfo {year} {1999})}\BibitemShut {NoStop}%
\bibitem [{\citenamefont {Nagaosa}\ \emph {et~al.}(2010)\citenamefont
  {Nagaosa}, \citenamefont {Sinova}, \citenamefont {Onoda}, \citenamefont
  {MacDonald},\ and\ \citenamefont {Ong}}]{nagaosa2010anomalous}%
  \BibitemOpen
  \bibfield  {author} {\bibinfo {author} {\bibfnamefont {N.}~\bibnamefont
  {Nagaosa}}, \bibinfo {author} {\bibfnamefont {J.}~\bibnamefont {Sinova}},
  \bibinfo {author} {\bibfnamefont {S.}~\bibnamefont {Onoda}}, \bibinfo
  {author} {\bibfnamefont {A.~H.}\ \bibnamefont {MacDonald}}, \ and\ \bibinfo
  {author} {\bibfnamefont {N.~P.}\ \bibnamefont {Ong}},\ }\href {\doibase
  10.1103/RevModPhys.82.1539} {\bibfield  {journal} {\bibinfo  {journal} {Rev.
  Mod. Phys.}\ }\textbf {\bibinfo {volume} {82}},\ \bibinfo {pages} {1539}
  (\bibinfo {year} {2010})}\BibitemShut {NoStop}%
\bibitem [{\citenamefont {Aharonov}\ and\ \citenamefont
  {Bohm}(1959)}]{aharonov1959significance}%
  \BibitemOpen
  \bibfield  {author} {\bibinfo {author} {\bibfnamefont {Y.}~\bibnamefont
  {Aharonov}}\ and\ \bibinfo {author} {\bibfnamefont {D.}~\bibnamefont
  {Bohm}},\ }\href {\doibase 10.1103/PhysRev.115.485} {\bibfield  {journal}
  {\bibinfo  {journal} {Phys. Rev.}\ }\textbf {\bibinfo {volume} {115}},\
  \bibinfo {pages} {485} (\bibinfo {year} {1959})}\BibitemShut {NoStop}%
\bibitem [{\citenamefont {Hasan}\ and\ \citenamefont
  {Kane}(2010)}]{hasan2010colloquium}%
  \BibitemOpen
  \bibfield  {author} {\bibinfo {author} {\bibfnamefont {M.~Z.}\ \bibnamefont
  {Hasan}}\ and\ \bibinfo {author} {\bibfnamefont {C.~L.}\ \bibnamefont
  {Kane}},\ }\href {\doibase 10.1103/RevModPhys.82.3045} {\bibfield  {journal}
  {\bibinfo  {journal} {Rev. Mod. Phys.}\ }\textbf {\bibinfo {volume} {82}},\
  \bibinfo {pages} {3045} (\bibinfo {year} {2010})}\BibitemShut {NoStop}%
\bibitem [{\citenamefont {Qi}\ and\ \citenamefont
  {Zhang}(2011)}]{qi2011topological}%
  \BibitemOpen
  \bibfield  {author} {\bibinfo {author} {\bibfnamefont {X.-L.}\ \bibnamefont
  {Qi}}\ and\ \bibinfo {author} {\bibfnamefont {S.-C.}\ \bibnamefont {Zhang}},\
  }\href {\doibase 10.1103/RevModPhys.83.1057} {\bibfield  {journal} {\bibinfo
  {journal} {Rev. Mod. Phys.}\ }\textbf {\bibinfo {volume} {83}},\ \bibinfo
  {pages} {1057} (\bibinfo {year} {2011})}\BibitemShut {NoStop}%
\bibitem [{\citenamefont {Zhang}\ \emph {et~al.}(2018)\citenamefont {Zhang},
  \citenamefont {Zhu}, \citenamefont {Zhao}, \citenamefont {Yan},\ and\
  \citenamefont {Zhu}}]{zhang2018topological}%
  \BibitemOpen
  \bibfield  {author} {\bibinfo {author} {\bibfnamefont {D.-W.}\ \bibnamefont
  {Zhang}}, \bibinfo {author} {\bibfnamefont {Y.-Q.}\ \bibnamefont {Zhu}},
  \bibinfo {author} {\bibfnamefont {Y.}~\bibnamefont {Zhao}}, \bibinfo {author}
  {\bibfnamefont {H.}~\bibnamefont {Yan}}, \ and\ \bibinfo {author}
  {\bibfnamefont {S.-L.}\ \bibnamefont {Zhu}},\ }\href {\doibase
  10.1080/00018732.2019.1594094} {\bibfield  {journal} {\bibinfo  {journal}
  {Adv. Phys.}\ }\textbf {\bibinfo {volume} {67}},\ \bibinfo {pages} {253}
  (\bibinfo {year} {2018})}\BibitemShut {NoStop}%
\bibitem [{\citenamefont {Julku}\ \emph {et~al.}(2016)\citenamefont {Julku},
  \citenamefont {Peotta}, \citenamefont {Vanhala}, \citenamefont {Kim},\ and\
  \citenamefont {T{\"o}rm{\"a}}}]{julku2016geometric}%
  \BibitemOpen
  \bibfield  {author} {\bibinfo {author} {\bibfnamefont {A.}~\bibnamefont
  {Julku}}, \bibinfo {author} {\bibfnamefont {S.}~\bibnamefont {Peotta}},
  \bibinfo {author} {\bibfnamefont {T.~I.}\ \bibnamefont {Vanhala}}, \bibinfo
  {author} {\bibfnamefont {D.-H.}\ \bibnamefont {Kim}}, \ and\ \bibinfo
  {author} {\bibfnamefont {P.}~\bibnamefont {T{\"o}rm{\"a}}},\ }\href {\doibase
  10.1103/PhysRevLett.117.045303} {\bibfield  {journal} {\bibinfo  {journal}
  {Phys. Rev. Lett,}\ }\textbf {\bibinfo {volume} {117}},\ \bibinfo {pages}
  {045303} (\bibinfo {year} {2016})}\BibitemShut {NoStop}%
\bibitem [{\citenamefont {He}\ \emph {et~al.}(2021)\citenamefont {He},
  \citenamefont {Ding},\ and\ \citenamefont {Zhu}}]{he2021geometry}%
  \BibitemOpen
  \bibfield  {author} {\bibinfo {author} {\bibfnamefont {P.}~\bibnamefont
  {He}}, \bibinfo {author} {\bibfnamefont {H.-T.}\ \bibnamefont {Ding}}, \ and\
  \bibinfo {author} {\bibfnamefont {S.-L.}\ \bibnamefont {Zhu}},\ }\href
  {\doibase 10.1103/PhysRevA.103.043329} {\bibfield  {journal} {\bibinfo
  {journal} {Phys. Rev. A}\ }\textbf {\bibinfo {volume} {103}},\ \bibinfo
  {pages} {043329} (\bibinfo {year} {2021})}\BibitemShut {NoStop}%
\bibitem [{\citenamefont {Roy}(2014)}]{roy2014band}%
  \BibitemOpen
  \bibfield  {author} {\bibinfo {author} {\bibfnamefont {R.}~\bibnamefont
  {Roy}},\ }\href {\doibase 10.1103/PhysRevB.90.165139} {\bibfield  {journal}
  {\bibinfo  {journal} {Phys. Rev. B}\ }\textbf {\bibinfo {volume} {90}},\
  \bibinfo {pages} {165139} (\bibinfo {year} {2014})}\BibitemShut {NoStop}%
\bibitem [{\citenamefont {Lim}\ \emph {et~al.}(2015)\citenamefont {Lim},
  \citenamefont {Fuchs},\ and\ \citenamefont {Montambaux}}]{lim2015geometry}%
  \BibitemOpen
  \bibfield  {author} {\bibinfo {author} {\bibfnamefont {L.-K.}\ \bibnamefont
  {Lim}}, \bibinfo {author} {\bibfnamefont {J.-N.}\ \bibnamefont {Fuchs}}, \
  and\ \bibinfo {author} {\bibfnamefont {G.}~\bibnamefont {Montambaux}},\
  }\href {\doibase 10.1103/PhysRevA.92.063627} {\bibfield  {journal} {\bibinfo
  {journal} {Phys. Rev. A}\ }\textbf {\bibinfo {volume} {92}},\ \bibinfo
  {pages} {063627} (\bibinfo {year} {2015})}\BibitemShut {NoStop}%
\bibitem [{\citenamefont {Palumbo}\ and\ \citenamefont
  {Goldman}(2018)}]{palumbo2018revealing}%
  \BibitemOpen
  \bibfield  {author} {\bibinfo {author} {\bibfnamefont {G.}~\bibnamefont
  {Palumbo}}\ and\ \bibinfo {author} {\bibfnamefont {N.}~\bibnamefont
  {Goldman}},\ }\href {\doibase 10.1103/PhysRevLett.121.170401} {\bibfield
  {journal} {\bibinfo  {journal} {Phys. Rev. Lett.}\ }\textbf {\bibinfo
  {volume} {121}},\ \bibinfo {pages} {170401} (\bibinfo {year}
  {2018})}\BibitemShut {NoStop}%
\bibitem [{\citenamefont {Tan}\ \emph {et~al.}(2019)\citenamefont {Tan},
  \citenamefont {Zhang}, \citenamefont {Yang}, \citenamefont {Chu},
  \citenamefont {Zhu}, \citenamefont {Li}, \citenamefont {Yang}, \citenamefont
  {Song}, \citenamefont {Han}, \citenamefont {Li}, \citenamefont {Dong},
  \citenamefont {Yu}, \citenamefont {Yan}, \citenamefont {Zhu},\ and\
  \citenamefont {Yu}}]{TanXinsheng2019}%
  \BibitemOpen
  \bibfield  {author} {\bibinfo {author} {\bibfnamefont {X.}~\bibnamefont
  {Tan}}, \bibinfo {author} {\bibfnamefont {D.-W.}\ \bibnamefont {Zhang}},
  \bibinfo {author} {\bibfnamefont {Z.}~\bibnamefont {Yang}}, \bibinfo {author}
  {\bibfnamefont {J.}~\bibnamefont {Chu}}, \bibinfo {author} {\bibfnamefont
  {Y.-Q.}\ \bibnamefont {Zhu}}, \bibinfo {author} {\bibfnamefont
  {D.}~\bibnamefont {Li}}, \bibinfo {author} {\bibfnamefont {X.}~\bibnamefont
  {Yang}}, \bibinfo {author} {\bibfnamefont {S.}~\bibnamefont {Song}}, \bibinfo
  {author} {\bibfnamefont {Z.}~\bibnamefont {Han}}, \bibinfo {author}
  {\bibfnamefont {Z.}~\bibnamefont {Li}}, \bibinfo {author} {\bibfnamefont
  {Y.}~\bibnamefont {Dong}}, \bibinfo {author} {\bibfnamefont {H.-F.}\
  \bibnamefont {Yu}}, \bibinfo {author} {\bibfnamefont {H.}~\bibnamefont
  {Yan}}, \bibinfo {author} {\bibfnamefont {S.-L.}\ \bibnamefont {Zhu}}, \ and\
  \bibinfo {author} {\bibfnamefont {Y.}~\bibnamefont {Yu}},\ }\href {\doibase
  10.1103/PhysRevLett.122.210401} {\bibfield  {journal} {\bibinfo  {journal}
  {Phys. Rev. Lett.}\ }\textbf {\bibinfo {volume} {122}},\ \bibinfo {pages}
  {210401} (\bibinfo {year} {2019})}\BibitemShut {NoStop}%
\bibitem [{\citenamefont {Tan}\ \emph {et~al.}(2021)\citenamefont {Tan},
  \citenamefont {Zhang}, \citenamefont {Zheng}, \citenamefont {Yang},
  \citenamefont {Song}, \citenamefont {Han}, \citenamefont {Dong},
  \citenamefont {Wang}, \citenamefont {Lan}, \citenamefont {Yan}, \citenamefont
  {Zhu},\ and\ \citenamefont {Yu}}]{TanXinsheng2021}%
  \BibitemOpen
  \bibfield  {author} {\bibinfo {author} {\bibfnamefont {X.}~\bibnamefont
  {Tan}}, \bibinfo {author} {\bibfnamefont {D.-W.}\ \bibnamefont {Zhang}},
  \bibinfo {author} {\bibfnamefont {W.}~\bibnamefont {Zheng}}, \bibinfo
  {author} {\bibfnamefont {X.}~\bibnamefont {Yang}}, \bibinfo {author}
  {\bibfnamefont {S.}~\bibnamefont {Song}}, \bibinfo {author} {\bibfnamefont
  {Z.}~\bibnamefont {Han}}, \bibinfo {author} {\bibfnamefont {Y.}~\bibnamefont
  {Dong}}, \bibinfo {author} {\bibfnamefont {Z.}~\bibnamefont {Wang}}, \bibinfo
  {author} {\bibfnamefont {D.}~\bibnamefont {Lan}}, \bibinfo {author}
  {\bibfnamefont {H.}~\bibnamefont {Yan}}, \bibinfo {author} {\bibfnamefont
  {S.-L.}\ \bibnamefont {Zhu}}, \ and\ \bibinfo {author} {\bibfnamefont
  {Y.}~\bibnamefont {Yu}},\ }\href {\doibase 10.1103/PhysRevLett.126.017702}
  {\bibfield  {journal} {\bibinfo  {journal} {Phys. Rev. Lett.}\ }\textbf
  {\bibinfo {volume} {126}},\ \bibinfo {pages} {017702} (\bibinfo {year}
  {2021})}\BibitemShut {NoStop}%
\bibitem [{\citenamefont {Liao}\ \emph {et~al.}(2021)\citenamefont {Liao},
  \citenamefont {Leblanc}, \citenamefont {Ren}, \citenamefont {Li},
  \citenamefont {Li}, \citenamefont {Solnyshkov}, \citenamefont {Malpuech},
  \citenamefont {Yao},\ and\ \citenamefont {Fu}}]{liao2021experimental}%
  \BibitemOpen
  \bibfield  {author} {\bibinfo {author} {\bibfnamefont {Q.}~\bibnamefont
  {Liao}}, \bibinfo {author} {\bibfnamefont {C.}~\bibnamefont {Leblanc}},
  \bibinfo {author} {\bibfnamefont {J.}~\bibnamefont {Ren}}, \bibinfo {author}
  {\bibfnamefont {F.}~\bibnamefont {Li}}, \bibinfo {author} {\bibfnamefont
  {Y.}~\bibnamefont {Li}}, \bibinfo {author} {\bibfnamefont {D.}~\bibnamefont
  {Solnyshkov}}, \bibinfo {author} {\bibfnamefont {G.}~\bibnamefont
  {Malpuech}}, \bibinfo {author} {\bibfnamefont {J.}~\bibnamefont {Yao}}, \
  and\ \bibinfo {author} {\bibfnamefont {H.}~\bibnamefont {Fu}},\ }\href
  {\doibase 10.1103/PhysRevLett.127.107402} {\bibfield  {journal} {\bibinfo
  {journal} {Phys. Rev. Lett.}\ }\textbf {\bibinfo {volume} {127}},\ \bibinfo
  {pages} {107402} (\bibinfo {year} {2021})}\BibitemShut {NoStop}%
\bibitem [{\citenamefont {Chen}\ \emph {et~al.}(2022)\citenamefont {Chen},
  \citenamefont {Li}, \citenamefont {Palumbo}, \citenamefont {Zhu},
  \citenamefont {Goldman},\ and\ \citenamefont
  {Cappellaro}}]{chen2022synthetic}%
  \BibitemOpen
  \bibfield  {author} {\bibinfo {author} {\bibfnamefont {M.}~\bibnamefont
  {Chen}}, \bibinfo {author} {\bibfnamefont {C.}~\bibnamefont {Li}}, \bibinfo
  {author} {\bibfnamefont {G.}~\bibnamefont {Palumbo}}, \bibinfo {author}
  {\bibfnamefont {Y.-Q.}\ \bibnamefont {Zhu}}, \bibinfo {author} {\bibfnamefont
  {N.}~\bibnamefont {Goldman}}, \ and\ \bibinfo {author} {\bibfnamefont
  {P.}~\bibnamefont {Cappellaro}},\ }\href {\doibase
  full/10.1126/science.abe6437} {\bibfield  {journal} {\bibinfo  {journal}
  {Science}\ }\textbf {\bibinfo {volume} {375}},\ \bibinfo {pages} {1017}
  (\bibinfo {year} {2022})}\BibitemShut {NoStop}%
\bibitem [{\citenamefont {Asteria}\ \emph {et~al.}(2019)\citenamefont
  {Asteria}, \citenamefont {Tran}, \citenamefont {Ozawa}, \citenamefont
  {Tarnowski}, \citenamefont {Rem}, \citenamefont {Fl{\"a}schner},
  \citenamefont {Sengstock}, \citenamefont {Goldman},\ and\ \citenamefont
  {Weitenberg}}]{asteria2019measuring}%
  \BibitemOpen
  \bibfield  {author} {\bibinfo {author} {\bibfnamefont {L.}~\bibnamefont
  {Asteria}}, \bibinfo {author} {\bibfnamefont {D.~T.}\ \bibnamefont {Tran}},
  \bibinfo {author} {\bibfnamefont {T.}~\bibnamefont {Ozawa}}, \bibinfo
  {author} {\bibfnamefont {M.}~\bibnamefont {Tarnowski}}, \bibinfo {author}
  {\bibfnamefont {B.~S.}\ \bibnamefont {Rem}}, \bibinfo {author} {\bibfnamefont
  {N.}~\bibnamefont {Fl{\"a}schner}}, \bibinfo {author} {\bibfnamefont
  {K.}~\bibnamefont {Sengstock}}, \bibinfo {author} {\bibfnamefont
  {N.}~\bibnamefont {Goldman}}, \ and\ \bibinfo {author} {\bibfnamefont
  {C.}~\bibnamefont {Weitenberg}},\ }\href {\doibase
  https://doi.org/10.1038/s41567-019-0417-8} {\bibfield  {journal} {\bibinfo
  {journal} {Nat. Phys.}\ }\textbf {\bibinfo {volume} {15}},\ \bibinfo {pages}
  {449} (\bibinfo {year} {2019})}\BibitemShut {NoStop}%
\bibitem [{\citenamefont {Yu}\ \emph {et~al.}(2022)\citenamefont {Yu},
  \citenamefont {Liu}, \citenamefont {Yang}, \citenamefont {Gong},
  \citenamefont {Cao}, \citenamefont {Zhang}, \citenamefont {Liu},
  \citenamefont {Heyl}, \citenamefont {Ozawa}, \citenamefont {Goldman} \emph
  {et~al.}}]{yu2022quantum}%
  \BibitemOpen
  \bibfield  {author} {\bibinfo {author} {\bibfnamefont {M.}~\bibnamefont
  {Yu}}, \bibinfo {author} {\bibfnamefont {Y.}~\bibnamefont {Liu}}, \bibinfo
  {author} {\bibfnamefont {P.}~\bibnamefont {Yang}}, \bibinfo {author}
  {\bibfnamefont {M.}~\bibnamefont {Gong}}, \bibinfo {author} {\bibfnamefont
  {Q.}~\bibnamefont {Cao}}, \bibinfo {author} {\bibfnamefont {S.}~\bibnamefont
  {Zhang}}, \bibinfo {author} {\bibfnamefont {H.}~\bibnamefont {Liu}}, \bibinfo
  {author} {\bibfnamefont {M.}~\bibnamefont {Heyl}}, \bibinfo {author}
  {\bibfnamefont {T.}~\bibnamefont {Ozawa}}, \bibinfo {author} {\bibfnamefont
  {N.}~\bibnamefont {Goldman}},  \emph {et~al.},\ }\href {\doibase
  https://doi.org/10.1038/s41534-022-00547-x} {\bibfield  {journal} {\bibinfo
  {journal} {npj Quantum Information}\ }\textbf {\bibinfo {volume} {8}},\
  \bibinfo {pages} {56} (\bibinfo {year} {2022})}\BibitemShut {NoStop}%
\bibitem [{\citenamefont {Zheng}\ \emph {et~al.}(2022)\citenamefont {Zheng},
  \citenamefont {Xu}, \citenamefont {Ma}, \citenamefont {Li}, \citenamefont
  {Dong}, \citenamefont {Zhang}, \citenamefont {Wang}, \citenamefont {Sun},
  \citenamefont {Wu}, \citenamefont {Zhao}, \citenamefont {Li}, \citenamefont
  {Lan}, \citenamefont {Tan},\ and\ \citenamefont {Yu}}]{zheng2022measuring}%
  \BibitemOpen
  \bibfield  {author} {\bibinfo {author} {\bibfnamefont {W.}~\bibnamefont
  {Zheng}}, \bibinfo {author} {\bibfnamefont {J.}~\bibnamefont {Xu}}, \bibinfo
  {author} {\bibfnamefont {Z.}~\bibnamefont {Ma}}, \bibinfo {author}
  {\bibfnamefont {Y.}~\bibnamefont {Li}}, \bibinfo {author} {\bibfnamefont
  {Y.}~\bibnamefont {Dong}}, \bibinfo {author} {\bibfnamefont {Y.}~\bibnamefont
  {Zhang}}, \bibinfo {author} {\bibfnamefont {X.}~\bibnamefont {Wang}},
  \bibinfo {author} {\bibfnamefont {G.}~\bibnamefont {Sun}}, \bibinfo {author}
  {\bibfnamefont {P.}~\bibnamefont {Wu}}, \bibinfo {author} {\bibfnamefont
  {J.}~\bibnamefont {Zhao}}, \bibinfo {author} {\bibfnamefont {S.}~\bibnamefont
  {Li}}, \bibinfo {author} {\bibfnamefont {D.}~\bibnamefont {Lan}}, \bibinfo
  {author} {\bibfnamefont {X.}~\bibnamefont {Tan}}, \ and\ \bibinfo {author}
  {\bibfnamefont {Y.}~\bibnamefont {Yu}},\ }\href {\doibase
  10.1088/0256-307X/39/10/100202} {\bibfield  {journal} {\bibinfo  {journal}
  {Chin. Phys. Lett.}\ }\textbf {\bibinfo {volume} {39}},\ \bibinfo {pages}
  {100202} (\bibinfo {year} {2022})}\BibitemShut {NoStop}%
\bibitem [{\citenamefont {Lysne}\ \emph {et~al.}(2023)\citenamefont {Lysne},
  \citenamefont {Sch\"uler},\ and\ \citenamefont {Werner}}]{lysne2023quantum}%
  \BibitemOpen
  \bibfield  {author} {\bibinfo {author} {\bibfnamefont {M.}~\bibnamefont
  {Lysne}}, \bibinfo {author} {\bibfnamefont {M.}~\bibnamefont {Sch\"uler}}, \
  and\ \bibinfo {author} {\bibfnamefont {P.}~\bibnamefont {Werner}},\ }\href
  {\doibase 10.1103/PhysRevLett.131.156901} {\bibfield  {journal} {\bibinfo
  {journal} {Phys. Rev. Lett.}\ }\textbf {\bibinfo {volume} {131}},\ \bibinfo
  {pages} {156901} (\bibinfo {year} {2023})}\BibitemShut {NoStop}%
\bibitem [{\citenamefont {Tan}\ \emph {et~al.}(2018)\citenamefont {Tan},
  \citenamefont {Zhang}, \citenamefont {Liu}, \citenamefont {Xue},
  \citenamefont {Yu}, \citenamefont {Zhu}, \citenamefont {Yan}, \citenamefont
  {Zhu},\ and\ \citenamefont {Yu}}]{tan2018topological}%
  \BibitemOpen
  \bibfield  {author} {\bibinfo {author} {\bibfnamefont {X.}~\bibnamefont
  {Tan}}, \bibinfo {author} {\bibfnamefont {D.-W.}\ \bibnamefont {Zhang}},
  \bibinfo {author} {\bibfnamefont {Q.}~\bibnamefont {Liu}}, \bibinfo {author}
  {\bibfnamefont {G.}~\bibnamefont {Xue}}, \bibinfo {author} {\bibfnamefont
  {H.-F.}\ \bibnamefont {Yu}}, \bibinfo {author} {\bibfnamefont {Y.-Q.}\
  \bibnamefont {Zhu}}, \bibinfo {author} {\bibfnamefont {H.}~\bibnamefont
  {Yan}}, \bibinfo {author} {\bibfnamefont {S.-L.}\ \bibnamefont {Zhu}}, \ and\
  \bibinfo {author} {\bibfnamefont {Y.}~\bibnamefont {Yu}},\ }\href {\doibase
  10.1103/PhysRevLett.120.130503} {\bibfield  {journal} {\bibinfo  {journal}
  {Phys. Rev. Lett.}\ }\textbf {\bibinfo {volume} {120}},\ \bibinfo {pages}
  {130503} (\bibinfo {year} {2018})}\BibitemShut {NoStop}%
\bibitem [{\citenamefont {Yu}\ \emph {et~al.}(2019)\citenamefont {Yu},
  \citenamefont {Yang}, \citenamefont {Gong}, \citenamefont {Cao},
  \citenamefont {Lu}, \citenamefont {Liu}, \citenamefont {Zhang}, \citenamefont
  {Plenio}, \citenamefont {Jelezko}, \citenamefont {Ozawa}, \citenamefont
  {Goldman},\ and\ \citenamefont {Cai}}]{yu2020experimental}%
  \BibitemOpen
  \bibfield  {author} {\bibinfo {author} {\bibfnamefont {M.}~\bibnamefont
  {Yu}}, \bibinfo {author} {\bibfnamefont {P.}~\bibnamefont {Yang}}, \bibinfo
  {author} {\bibfnamefont {M.}~\bibnamefont {Gong}}, \bibinfo {author}
  {\bibfnamefont {Q.}~\bibnamefont {Cao}}, \bibinfo {author} {\bibfnamefont
  {Q.}~\bibnamefont {Lu}}, \bibinfo {author} {\bibfnamefont {H.}~\bibnamefont
  {Liu}}, \bibinfo {author} {\bibfnamefont {S.}~\bibnamefont {Zhang}}, \bibinfo
  {author} {\bibfnamefont {M.~B.}\ \bibnamefont {Plenio}}, \bibinfo {author}
  {\bibfnamefont {F.}~\bibnamefont {Jelezko}}, \bibinfo {author} {\bibfnamefont
  {T.}~\bibnamefont {Ozawa}}, \bibinfo {author} {\bibfnamefont
  {N.}~\bibnamefont {Goldman}}, \ and\ \bibinfo {author} {\bibfnamefont
  {J.}~\bibnamefont {Cai}},\ }\href {\doibase 10.1093/nsr/nwz193} {\bibfield
  {journal} {\bibinfo  {journal} {Natl. Sci. Rev.}\ }\textbf {\bibinfo {volume}
  {7}},\ \bibinfo {pages} {254} (\bibinfo {year} {2019})}\BibitemShut {NoStop}%
\bibitem [{\citenamefont {Yu}\ \emph {et~al.}(2024)\citenamefont {Yu},
  \citenamefont {Li}, \citenamefont {Chu}, \citenamefont {Mera}, \citenamefont
  {Unal}, \citenamefont {Yang}, \citenamefont {Liu}, \citenamefont {Goldman},\
  and\ \citenamefont {Cai}}]{yu2022experimental}%
  \BibitemOpen
  \bibfield  {author} {\bibinfo {author} {\bibfnamefont {M.}~\bibnamefont
  {Yu}}, \bibinfo {author} {\bibfnamefont {X.}~\bibnamefont {Li}}, \bibinfo
  {author} {\bibfnamefont {Y.}~\bibnamefont {Chu}}, \bibinfo {author}
  {\bibfnamefont {B.}~\bibnamefont {Mera}}, \bibinfo {author} {\bibfnamefont
  {F.~N.}\ \bibnamefont {Unal}}, \bibinfo {author} {\bibfnamefont
  {P.}~\bibnamefont {Yang}}, \bibinfo {author} {\bibfnamefont {Y.}~\bibnamefont
  {Liu}}, \bibinfo {author} {\bibfnamefont {N.}~\bibnamefont {Goldman}}, \ and\
  \bibinfo {author} {\bibfnamefont {J.}~\bibnamefont {Cai}},\ }\href {\doibase
  10.1093/nsr/nwae065} {\bibfield  {journal} {\bibinfo  {journal} {Natl. Sci.
  Rev.}\ ,\ \bibinfo {pages} {nwae065}} (\bibinfo {year} {2024})}\BibitemShut
  {NoStop}%
\bibitem [{\citenamefont {Gianfrate}\ \emph {et~al.}(2020)\citenamefont
  {Gianfrate}, \citenamefont {Bleu}, \citenamefont {Dominici}, \citenamefont
  {Ardizzone}, \citenamefont {De~Giorgi}, \citenamefont {Ballarini},
  \citenamefont {Lerario}, \citenamefont {West}, \citenamefont {Pfeiffer},
  \citenamefont {Solnyshkov} \emph {et~al.}}]{gianfrate2020measurement}%
  \BibitemOpen
  \bibfield  {author} {\bibinfo {author} {\bibfnamefont {A.}~\bibnamefont
  {Gianfrate}}, \bibinfo {author} {\bibfnamefont {O.}~\bibnamefont {Bleu}},
  \bibinfo {author} {\bibfnamefont {L.}~\bibnamefont {Dominici}}, \bibinfo
  {author} {\bibfnamefont {V.}~\bibnamefont {Ardizzone}}, \bibinfo {author}
  {\bibfnamefont {M.}~\bibnamefont {De~Giorgi}}, \bibinfo {author}
  {\bibfnamefont {D.}~\bibnamefont {Ballarini}}, \bibinfo {author}
  {\bibfnamefont {G.}~\bibnamefont {Lerario}}, \bibinfo {author} {\bibfnamefont
  {K.}~\bibnamefont {West}}, \bibinfo {author} {\bibfnamefont {L.}~\bibnamefont
  {Pfeiffer}}, \bibinfo {author} {\bibfnamefont {D.}~\bibnamefont
  {Solnyshkov}},  \emph {et~al.},\ }\href {\doibase
  https://doi.org/10.1038/s41586-020-1989-2} {\bibfield  {journal} {\bibinfo
  {journal} {Nature}\ }\textbf {\bibinfo {volume} {578}},\ \bibinfo {pages}
  {381} (\bibinfo {year} {2020})}\BibitemShut {NoStop}%
\bibitem [{\citenamefont {Yi}\ \emph {et~al.}(2023)\citenamefont {Yi},
  \citenamefont {Yu}, \citenamefont {Yuan}, \citenamefont {Jiao}, \citenamefont
  {Yang}, \citenamefont {Jiang}, \citenamefont {Zhang}, \citenamefont {Chen},\
  and\ \citenamefont {Pan}}]{yi2023extracting}%
  \BibitemOpen
  \bibfield  {author} {\bibinfo {author} {\bibfnamefont {C.-R.}\ \bibnamefont
  {Yi}}, \bibinfo {author} {\bibfnamefont {J.}~\bibnamefont {Yu}}, \bibinfo
  {author} {\bibfnamefont {H.}~\bibnamefont {Yuan}}, \bibinfo {author}
  {\bibfnamefont {R.-H.}\ \bibnamefont {Jiao}}, \bibinfo {author}
  {\bibfnamefont {Y.-M.}\ \bibnamefont {Yang}}, \bibinfo {author}
  {\bibfnamefont {X.}~\bibnamefont {Jiang}}, \bibinfo {author} {\bibfnamefont
  {J.-Y.}\ \bibnamefont {Zhang}}, \bibinfo {author} {\bibfnamefont
  {S.}~\bibnamefont {Chen}}, \ and\ \bibinfo {author} {\bibfnamefont {J.-W.}\
  \bibnamefont {Pan}},\ }\href {\doibase 10.1103/PhysRevResearch.5.L032016}
  {\bibfield  {journal} {\bibinfo  {journal} {Phys. Rev. Res.}\ }\textbf
  {\bibinfo {volume} {5}},\ \bibinfo {pages} {L032016} (\bibinfo {year}
  {2023})}\BibitemShut {NoStop}%
\bibitem [{\citenamefont {Cuerda}\ \emph {et~al.}()\citenamefont {Cuerda},
  \citenamefont {Taskinen}, \citenamefont {Källman}, \citenamefont {Grabitz},\
  and\ \citenamefont {Törmä}}]{cuerda2024observation}%
  \BibitemOpen
  \bibfield  {author} {\bibinfo {author} {\bibfnamefont {J.}~\bibnamefont
  {Cuerda}}, \bibinfo {author} {\bibfnamefont {J.~M.}\ \bibnamefont
  {Taskinen}}, \bibinfo {author} {\bibfnamefont {N.}~\bibnamefont {Källman}},
  \bibinfo {author} {\bibfnamefont {L.}~\bibnamefont {Grabitz}}, \ and\
  \bibinfo {author} {\bibfnamefont {P.}~\bibnamefont {Törmä}},\ }\href@noop
  {} {}\Eprint {http://arxiv.org/abs/2305.13174} {arXiv:2305.13174}
  \BibitemShut {NoStop}%
\bibitem [{\citenamefont {Solnyshkov}\ \emph {et~al.}(2021)\citenamefont
  {Solnyshkov}, \citenamefont {Leblanc}, \citenamefont {Bessonart},
  \citenamefont {Nalitov}, \citenamefont {Ren}, \citenamefont {Liao},
  \citenamefont {Li},\ and\ \citenamefont {Malpuech}}]{SolnyshkovD.D.2021}%
  \BibitemOpen
  \bibfield  {author} {\bibinfo {author} {\bibfnamefont {D.~D.}\ \bibnamefont
  {Solnyshkov}}, \bibinfo {author} {\bibfnamefont {C.}~\bibnamefont {Leblanc}},
  \bibinfo {author} {\bibfnamefont {L.}~\bibnamefont {Bessonart}}, \bibinfo
  {author} {\bibfnamefont {A.}~\bibnamefont {Nalitov}}, \bibinfo {author}
  {\bibfnamefont {J.}~\bibnamefont {Ren}}, \bibinfo {author} {\bibfnamefont
  {Q.}~\bibnamefont {Liao}}, \bibinfo {author} {\bibfnamefont {F.}~\bibnamefont
  {Li}}, \ and\ \bibinfo {author} {\bibfnamefont {G.}~\bibnamefont
  {Malpuech}},\ }\href {\doibase 10.1103/PhysRevB.103.125302} {\bibfield
  {journal} {\bibinfo  {journal} {Phys. Rev. B}\ }\textbf {\bibinfo {volume}
  {103}},\ \bibinfo {pages} {125302} (\bibinfo {year} {2021})}\BibitemShut
  {NoStop}%
\bibitem [{\citenamefont {Campos~Venuti}\ and\ \citenamefont
  {Zanardi}(2007)}]{CamposVenuti2007}%
  \BibitemOpen
  \bibfield  {author} {\bibinfo {author} {\bibfnamefont {L.}~\bibnamefont
  {Campos~Venuti}}\ and\ \bibinfo {author} {\bibfnamefont {P.}~\bibnamefont
  {Zanardi}},\ }\href {\doibase 10.1103/PhysRevLett.99.095701} {\bibfield
  {journal} {\bibinfo  {journal} {Phys. Rev. Lett.}\ }\textbf {\bibinfo
  {volume} {99}},\ \bibinfo {pages} {095701} (\bibinfo {year}
  {2007})}\BibitemShut {NoStop}%
\bibitem [{\citenamefont {Zanardi}\ \emph
  {et~al.}(2007{\natexlab{b}})\citenamefont {Zanardi}, \citenamefont {Giorda},\
  and\ \citenamefont {Cozzini}}]{ZanardiPaolo2007}%
  \BibitemOpen
  \bibfield  {author} {\bibinfo {author} {\bibfnamefont {P.}~\bibnamefont
  {Zanardi}}, \bibinfo {author} {\bibfnamefont {P.}~\bibnamefont {Giorda}}, \
  and\ \bibinfo {author} {\bibfnamefont {M.}~\bibnamefont {Cozzini}},\ }\href
  {\doibase 10.1103/PhysRevLett.99.100603} {\bibfield  {journal} {\bibinfo
  {journal} {Phys. Rev. Lett.}\ }\textbf {\bibinfo {volume} {99}},\ \bibinfo
  {pages} {100603} (\bibinfo {year} {2007}{\natexlab{b}})}\BibitemShut
  {NoStop}%
\bibitem [{\citenamefont {Zhang}\ \emph {et~al.}({\natexlab{a}})\citenamefont
  {Zhang}, \citenamefont {Lv}, \citenamefont {Chen}, \citenamefont {Yu},
  \citenamefont {Wu}, \citenamefont {Yang},\ and\ \citenamefont
  {Zheng}}]{zhang2023}%
  \BibitemOpen
  \bibfield  {author} {\bibinfo {author} {\bibfnamefont {H.-L.}\ \bibnamefont
  {Zhang}}, \bibinfo {author} {\bibfnamefont {J.-H.}\ \bibnamefont {Lv}},
  \bibinfo {author} {\bibfnamefont {K.}~\bibnamefont {Chen}}, \bibinfo {author}
  {\bibfnamefont {X.-J.}\ \bibnamefont {Yu}}, \bibinfo {author} {\bibfnamefont
  {F.}~\bibnamefont {Wu}}, \bibinfo {author} {\bibfnamefont {Z.-B.}\
  \bibnamefont {Yang}}, \ and\ \bibinfo {author} {\bibfnamefont {S.-B.}\
  \bibnamefont {Zheng}},\ }\href@noop {} {} ({\natexlab{a}}),\ \Eprint
  {http://arxiv.org/abs/2312.14414} {arXiv:2312.14414} \BibitemShut {NoStop}%
\bibitem [{\citenamefont {Abasto}\ \emph {et~al.}(2008)\citenamefont {Abasto},
  \citenamefont {Hamma},\ and\ \citenamefont {Zanardi}}]{Abasto2008}%
  \BibitemOpen
  \bibfield  {author} {\bibinfo {author} {\bibfnamefont {D.~F.}\ \bibnamefont
  {Abasto}}, \bibinfo {author} {\bibfnamefont {A.}~\bibnamefont {Hamma}}, \
  and\ \bibinfo {author} {\bibfnamefont {P.}~\bibnamefont {Zanardi}},\ }\href
  {\doibase 10.1103/PhysRevA.78.010301} {\bibfield  {journal} {\bibinfo
  {journal} {Phys. Rev. A}\ }\textbf {\bibinfo {volume} {78}},\ \bibinfo
  {pages} {010301} (\bibinfo {year} {2008})}\BibitemShut {NoStop}%
\bibitem [{\citenamefont {Het{\'e}nyi}\ and\ \citenamefont
  {L{\'e}vay}(2023)}]{hetenyi2023fluctuations}%
  \BibitemOpen
  \bibfield  {author} {\bibinfo {author} {\bibfnamefont {B.}~\bibnamefont
  {Het{\'e}nyi}}\ and\ \bibinfo {author} {\bibfnamefont {P.}~\bibnamefont
  {L{\'e}vay}},\ }\href {\doibase 10.1103/PhysRevA.108.032218} {\bibfield
  {journal} {\bibinfo  {journal} {Phys. Rev. A}\ }\textbf {\bibinfo {volume}
  {108}},\ \bibinfo {pages} {032218} (\bibinfo {year} {2023})}\BibitemShut
  {NoStop}%
\bibitem [{\citenamefont {Albuquerque}\ \emph {et~al.}(2010)\citenamefont
  {Albuquerque}, \citenamefont {Alet}, \citenamefont {Sire},\ and\
  \citenamefont {Capponi}}]{Albuquerque2010}%
  \BibitemOpen
  \bibfield  {author} {\bibinfo {author} {\bibfnamefont {A.~F.}\ \bibnamefont
  {Albuquerque}}, \bibinfo {author} {\bibfnamefont {F.}~\bibnamefont {Alet}},
  \bibinfo {author} {\bibfnamefont {C.}~\bibnamefont {Sire}}, \ and\ \bibinfo
  {author} {\bibfnamefont {S.}~\bibnamefont {Capponi}},\ }\href {\doibase
  10.1103/PhysRevB.81.064418} {\bibfield  {journal} {\bibinfo  {journal} {Phys.
  Rev. B}\ }\textbf {\bibinfo {volume} {81}},\ \bibinfo {pages} {064418}
  (\bibinfo {year} {2010})}\BibitemShut {NoStop}%
\bibitem [{\citenamefont {Dey}\ \emph {et~al.}(2012)\citenamefont {Dey},
  \citenamefont {Mahapatra}, \citenamefont {Roy},\ and\ \citenamefont
  {Sarkar}}]{Anshuman2012}%
  \BibitemOpen
  \bibfield  {author} {\bibinfo {author} {\bibfnamefont {A.}~\bibnamefont
  {Dey}}, \bibinfo {author} {\bibfnamefont {S.}~\bibnamefont {Mahapatra}},
  \bibinfo {author} {\bibfnamefont {P.}~\bibnamefont {Roy}}, \ and\ \bibinfo
  {author} {\bibfnamefont {T.}~\bibnamefont {Sarkar}},\ }\href {\doibase
  10.1103/PhysRevE.86.031137} {\bibfield  {journal} {\bibinfo  {journal} {Phys.
  Rev. E}\ }\textbf {\bibinfo {volume} {86}},\ \bibinfo {pages} {031137}
  (\bibinfo {year} {2012})}\BibitemShut {NoStop}%
\bibitem [{\citenamefont {Zhang}\ \emph {et~al.}({\natexlab{b}})\citenamefont
  {Zhang}, \citenamefont {Lv}, \citenamefont {Chen}, \citenamefont {Yu},
  \citenamefont {Wu}, \citenamefont {Yang},\ and\ \citenamefont
  {Zheng}}]{zhang2023critical}%
  \BibitemOpen
  \bibfield  {author} {\bibinfo {author} {\bibfnamefont {H.-L.}\ \bibnamefont
  {Zhang}}, \bibinfo {author} {\bibfnamefont {J.-H.}\ \bibnamefont {Lv}},
  \bibinfo {author} {\bibfnamefont {K.}~\bibnamefont {Chen}}, \bibinfo {author}
  {\bibfnamefont {X.-J.}\ \bibnamefont {Yu}}, \bibinfo {author} {\bibfnamefont
  {F.}~\bibnamefont {Wu}}, \bibinfo {author} {\bibfnamefont {Z.-B.}\
  \bibnamefont {Yang}}, \ and\ \bibinfo {author} {\bibfnamefont {S.-B.}\
  \bibnamefont {Zheng}},\ }\href@noop {} {} ({\natexlab{b}}),\ \Eprint
  {http://arxiv.org/abs/2312.14414} {arXiv:2312.14414 [quant-ph]} \BibitemShut
  {NoStop}%
\bibitem [{\citenamefont {Ashida}\ \emph {et~al.}(2020)\citenamefont {Ashida},
  \citenamefont {Gong},\ and\ \citenamefont {Ueda}}]{ashida2020non}%
  \BibitemOpen
  \bibfield  {author} {\bibinfo {author} {\bibfnamefont {Y.}~\bibnamefont
  {Ashida}}, \bibinfo {author} {\bibfnamefont {Z.}~\bibnamefont {Gong}}, \ and\
  \bibinfo {author} {\bibfnamefont {M.}~\bibnamefont {Ueda}},\ }\href {\doibase
  10.1080/00018732.2021.1876991} {\bibfield  {journal} {\bibinfo  {journal}
  {Advances in Physics}\ }\textbf {\bibinfo {volume} {69}},\ \bibinfo {pages}
  {249} (\bibinfo {year} {2020})}\BibitemShut {NoStop}%
\bibitem [{\citenamefont {Hatano}\ and\ \citenamefont
  {Nelson}(1996)}]{HatanoNaomichi1996}%
  \BibitemOpen
  \bibfield  {author} {\bibinfo {author} {\bibfnamefont {N.}~\bibnamefont
  {Hatano}}\ and\ \bibinfo {author} {\bibfnamefont {D.~R.}\ \bibnamefont
  {Nelson}},\ }\href {\doibase 10.1103/PhysRevLett.77.570} {\bibfield
  {journal} {\bibinfo  {journal} {Phys. Rev. Lett.}\ }\textbf {\bibinfo
  {volume} {77}},\ \bibinfo {pages} {570} (\bibinfo {year} {1996})}\BibitemShut
  {NoStop}%
\bibitem [{\citenamefont {Hatano}\ and\ \citenamefont
  {Nelson}(1997)}]{HatanoNaomichi1997}%
  \BibitemOpen
  \bibfield  {author} {\bibinfo {author} {\bibfnamefont {N.}~\bibnamefont
  {Hatano}}\ and\ \bibinfo {author} {\bibfnamefont {D.~R.}\ \bibnamefont
  {Nelson}},\ }\href {\doibase 10.1103/PhysRevB.56.8651} {\bibfield  {journal}
  {\bibinfo  {journal} {Phys. Rev. B}\ }\textbf {\bibinfo {volume} {56}},\
  \bibinfo {pages} {8651} (\bibinfo {year} {1997})}\BibitemShut {NoStop}%
\bibitem [{\citenamefont {Bender}\ and\ \citenamefont
  {Boettcher}(1998)}]{BenderCarlM.1998}%
  \BibitemOpen
  \bibfield  {author} {\bibinfo {author} {\bibfnamefont {C.~M.}\ \bibnamefont
  {Bender}}\ and\ \bibinfo {author} {\bibfnamefont {S.}~\bibnamefont
  {Boettcher}},\ }\href {\doibase 10.1103/PhysRevLett.80.5243} {\bibfield
  {journal} {\bibinfo  {journal} {Phys. Rev. Lett.}\ }\textbf {\bibinfo
  {volume} {80}},\ \bibinfo {pages} {5243} (\bibinfo {year}
  {1998})}\BibitemShut {NoStop}%
\bibitem [{\citenamefont {Liu}\ \emph {et~al.}(2019)\citenamefont {Liu},
  \citenamefont {Zhang}, \citenamefont {Ai}, \citenamefont {Gong},
  \citenamefont {Kawabata}, \citenamefont {Ueda},\ and\ \citenamefont
  {Nori}}]{LiuTao2019}%
  \BibitemOpen
  \bibfield  {author} {\bibinfo {author} {\bibfnamefont {T.}~\bibnamefont
  {Liu}}, \bibinfo {author} {\bibfnamefont {Y.-R.}\ \bibnamefont {Zhang}},
  \bibinfo {author} {\bibfnamefont {Q.}~\bibnamefont {Ai}}, \bibinfo {author}
  {\bibfnamefont {Z.}~\bibnamefont {Gong}}, \bibinfo {author} {\bibfnamefont
  {K.}~\bibnamefont {Kawabata}}, \bibinfo {author} {\bibfnamefont
  {M.}~\bibnamefont {Ueda}}, \ and\ \bibinfo {author} {\bibfnamefont
  {F.}~\bibnamefont {Nori}},\ }\href {\doibase 10.1103/PhysRevLett.122.076801}
  {\bibfield  {journal} {\bibinfo  {journal} {Phys. Rev. Lett.}\ }\textbf
  {\bibinfo {volume} {122}},\ \bibinfo {pages} {076801} (\bibinfo {year}
  {2019})}\BibitemShut {NoStop}%
\bibitem [{\citenamefont {Kawabata}\ \emph {et~al.}(2019)\citenamefont
  {Kawabata}, \citenamefont {Higashikawa}, \citenamefont {Gong}, \citenamefont
  {Ashida},\ and\ \citenamefont {Ueda}}]{kawabata2019}%
  \BibitemOpen
  \bibfield  {author} {\bibinfo {author} {\bibfnamefont {K.}~\bibnamefont
  {Kawabata}}, \bibinfo {author} {\bibfnamefont {S.}~\bibnamefont
  {Higashikawa}}, \bibinfo {author} {\bibfnamefont {Z.}~\bibnamefont {Gong}},
  \bibinfo {author} {\bibfnamefont {Y.}~\bibnamefont {Ashida}}, \ and\ \bibinfo
  {author} {\bibfnamefont {M.}~\bibnamefont {Ueda}},\ }\href {\doibase
  10.1038/s41467-018-08254-y} {\bibfield  {journal} {\bibinfo  {journal} {Nat.
  Commun.}\ }\textbf {\bibinfo {volume} {10}},\ \bibinfo {pages} {297}
  (\bibinfo {year} {2019})}\BibitemShut {NoStop}%
\bibitem [{\citenamefont {Chen}\ and\ \citenamefont {Zhai}(2018)}]{Chen2018}%
  \BibitemOpen
  \bibfield  {author} {\bibinfo {author} {\bibfnamefont {Y.}~\bibnamefont
  {Chen}}\ and\ \bibinfo {author} {\bibfnamefont {H.}~\bibnamefont {Zhai}},\
  }\href {\doibase 10.1103/PhysRevB.98.245130} {\bibfield  {journal} {\bibinfo
  {journal} {Phys. Rev. B}\ }\textbf {\bibinfo {volume} {98}},\ \bibinfo
  {pages} {245130} (\bibinfo {year} {2018})}\BibitemShut {NoStop}%
\bibitem [{\citenamefont {Zhang}\ \emph
  {et~al.}(2020{\natexlab{a}})\citenamefont {Zhang}, \citenamefont {Chen},
  \citenamefont {Zhang}, \citenamefont {Lang}, \citenamefont {Li},\ and\
  \citenamefont {Zhu}}]{DWZhang2020}%
  \BibitemOpen
  \bibfield  {author} {\bibinfo {author} {\bibfnamefont {D.-W.}\ \bibnamefont
  {Zhang}}, \bibinfo {author} {\bibfnamefont {Y.-L.}\ \bibnamefont {Chen}},
  \bibinfo {author} {\bibfnamefont {G.-Q.}\ \bibnamefont {Zhang}}, \bibinfo
  {author} {\bibfnamefont {L.-J.}\ \bibnamefont {Lang}}, \bibinfo {author}
  {\bibfnamefont {Z.}~\bibnamefont {Li}}, \ and\ \bibinfo {author}
  {\bibfnamefont {S.-L.}\ \bibnamefont {Zhu}},\ }\href {\doibase
  10.1103/PhysRevB.101.235150} {\bibfield  {journal} {\bibinfo  {journal}
  {Phys. Rev. B}\ }\textbf {\bibinfo {volume} {101}},\ \bibinfo {pages}
  {235150} (\bibinfo {year} {2020}{\natexlab{a}})}\BibitemShut {NoStop}%
\bibitem [{\citenamefont {Hu}\ \emph {et~al.}()\citenamefont {Hu},
  \citenamefont {Ostrovskaya},\ and\ \citenamefont {Estrecho}}]{hu2023}%
  \BibitemOpen
  \bibfield  {author} {\bibinfo {author} {\bibfnamefont {Y.~M.~R.}\
  \bibnamefont {Hu}}, \bibinfo {author} {\bibfnamefont {E.~A.}\ \bibnamefont
  {Ostrovskaya}}, \ and\ \bibinfo {author} {\bibfnamefont {E.}~\bibnamefont
  {Estrecho}},\ }\href@noop {} {}\Eprint {http://arxiv.org/abs/2306.00351}
  {arXiv:2306.00351} \BibitemShut {NoStop}%
\bibitem [{\citenamefont {Tu}\ \emph {et~al.}(2023)\citenamefont {Tu},
  \citenamefont {Jang}, \citenamefont {Chang},\ and\ \citenamefont
  {Tzeng}}]{Tu2023}%
  \BibitemOpen
  \bibfield  {author} {\bibinfo {author} {\bibfnamefont {Y.-T.}\ \bibnamefont
  {Tu}}, \bibinfo {author} {\bibfnamefont {I.}~\bibnamefont {Jang}}, \bibinfo
  {author} {\bibfnamefont {P.-Y.}\ \bibnamefont {Chang}}, \ and\ \bibinfo
  {author} {\bibfnamefont {Y.-C.}\ \bibnamefont {Tzeng}},\ }\href {\doibase
  10.22331/q-2023-03-23-960} {\bibfield  {journal} {\bibinfo  {journal}
  {{Quantum}}\ }\textbf {\bibinfo {volume} {7}},\ \bibinfo {pages} {960}
  (\bibinfo {year} {2023})}\BibitemShut {NoStop}%
\bibitem [{\citenamefont {Tzeng}\ \emph {et~al.}(2021)\citenamefont {Tzeng},
  \citenamefont {Ju}, \citenamefont {Chen},\ and\ \citenamefont
  {Huang}}]{TzengYu-Chin2021}%
  \BibitemOpen
  \bibfield  {author} {\bibinfo {author} {\bibfnamefont {Y.-C.}\ \bibnamefont
  {Tzeng}}, \bibinfo {author} {\bibfnamefont {C.-Y.}\ \bibnamefont {Ju}},
  \bibinfo {author} {\bibfnamefont {G.-Y.}\ \bibnamefont {Chen}}, \ and\
  \bibinfo {author} {\bibfnamefont {W.-M.}\ \bibnamefont {Huang}},\ }\href
  {\doibase 10.1103/PhysRevResearch.3.013015} {\bibfield  {journal} {\bibinfo
  {journal} {Phys. Rev. Res.}\ }\textbf {\bibinfo {volume} {3}},\ \bibinfo
  {pages} {013015} (\bibinfo {year} {2021})}\BibitemShut {NoStop}%
\bibitem [{\citenamefont {Sticlet}\ \emph {et~al.}(2023)\citenamefont
  {Sticlet}, \citenamefont {Moca},\ and\ \citenamefont {D\'ora}}]{Sticlet2023}%
  \BibitemOpen
  \bibfield  {author} {\bibinfo {author} {\bibfnamefont {D.}~\bibnamefont
  {Sticlet}}, \bibinfo {author} {\bibfnamefont {C.~P.}\ \bibnamefont {Moca}}, \
  and\ \bibinfo {author} {\bibfnamefont {B.}~\bibnamefont {D\'ora}},\ }\href
  {\doibase 10.1103/PhysRevB.108.075133} {\bibfield  {journal} {\bibinfo
  {journal} {Phys. Rev. B}\ }\textbf {\bibinfo {volume} {108}},\ \bibinfo
  {pages} {075133} (\bibinfo {year} {2023})}\BibitemShut {NoStop}%
\bibitem [{\citenamefont {D\'ora}\ \emph {et~al.}(2022)\citenamefont {D\'ora},
  \citenamefont {Sticlet},\ and\ \citenamefont {Moca}}]{ora2022}%
  \BibitemOpen
  \bibfield  {author} {\bibinfo {author} {\bibfnamefont {B.}~\bibnamefont
  {D\'ora}}, \bibinfo {author} {\bibfnamefont {D.}~\bibnamefont {Sticlet}}, \
  and\ \bibinfo {author} {\bibfnamefont {C.~P.}\ \bibnamefont {Moca}},\ }\href
  {\doibase 10.1103/PhysRevLett.128.146804} {\bibfield  {journal} {\bibinfo
  {journal} {Phys. Rev. Lett.}\ }\textbf {\bibinfo {volume} {128}},\ \bibinfo
  {pages} {146804} (\bibinfo {year} {2022})}\BibitemShut {NoStop}%
\bibitem [{\citenamefont {Yao}\ and\ \citenamefont {Wang}(2018)}]{Yao2018}%
  \BibitemOpen
  \bibfield  {author} {\bibinfo {author} {\bibfnamefont {S.-Y.}\ \bibnamefont
  {Yao}}\ and\ \bibinfo {author} {\bibfnamefont {Z.}~\bibnamefont {Wang}},\
  }\href {\doibase 10.1103/PhysRevLett.121.086803} {\bibfield  {journal}
  {\bibinfo  {journal} {Phys. Rev. Lett.}\ }\textbf {\bibinfo {volume} {121}},\
  \bibinfo {pages} {086803} (\bibinfo {year} {2018})}\BibitemShut {NoStop}%
\bibitem [{\citenamefont {Jiang}\ \emph {et~al.}(2019)\citenamefont {Jiang},
  \citenamefont {Lang}, \citenamefont {Yang}, \citenamefont {Zhu},\ and\
  \citenamefont {Chen}}]{JiangHui2019}%
  \BibitemOpen
  \bibfield  {author} {\bibinfo {author} {\bibfnamefont {H.}~\bibnamefont
  {Jiang}}, \bibinfo {author} {\bibfnamefont {L.-J.}\ \bibnamefont {Lang}},
  \bibinfo {author} {\bibfnamefont {C.}~\bibnamefont {Yang}}, \bibinfo {author}
  {\bibfnamefont {S.-L.}\ \bibnamefont {Zhu}}, \ and\ \bibinfo {author}
  {\bibfnamefont {S.}~\bibnamefont {Chen}},\ }\href {\doibase
  10.1103/PhysRevB.100.054301} {\bibfield  {journal} {\bibinfo  {journal}
  {Phys. Rev. B}\ }\textbf {\bibinfo {volume} {100}},\ \bibinfo {pages}
  {054301} (\bibinfo {year} {2019})}\BibitemShut {NoStop}%
\bibitem [{\citenamefont {Song}\ \emph {et~al.}(2019)\citenamefont {Song},
  \citenamefont {Yao},\ and\ \citenamefont {Wang}}]{Songfei2019}%
  \BibitemOpen
  \bibfield  {author} {\bibinfo {author} {\bibfnamefont {F.}~\bibnamefont
  {Song}}, \bibinfo {author} {\bibfnamefont {S.}~\bibnamefont {Yao}}, \ and\
  \bibinfo {author} {\bibfnamefont {Z.}~\bibnamefont {Wang}},\ }\href {\doibase
  10.1103/PhysRevLett.123.246801} {\bibfield  {journal} {\bibinfo  {journal}
  {Phys. Rev. Lett.}\ }\textbf {\bibinfo {volume} {123}},\ \bibinfo {pages}
  {246801} (\bibinfo {year} {2019})}\BibitemShut {NoStop}%
\bibitem [{\citenamefont {Tang}\ \emph {et~al.}(2020)\citenamefont {Tang},
  \citenamefont {Zhang}, \citenamefont {Zhang},\ and\ \citenamefont
  {Zhang}}]{LZTang2020}%
  \BibitemOpen
  \bibfield  {author} {\bibinfo {author} {\bibfnamefont {L.-Z.}\ \bibnamefont
  {Tang}}, \bibinfo {author} {\bibfnamefont {L.-F.}\ \bibnamefont {Zhang}},
  \bibinfo {author} {\bibfnamefont {G.-Q.}\ \bibnamefont {Zhang}}, \ and\
  \bibinfo {author} {\bibfnamefont {D.-W.}\ \bibnamefont {Zhang}},\ }\href
  {\doibase 10.1103/PhysRevA.101.063612} {\bibfield  {journal} {\bibinfo
  {journal} {Phys. Rev. A}\ }\textbf {\bibinfo {volume} {101}},\ \bibinfo
  {pages} {063612} (\bibinfo {year} {2020})}\BibitemShut {NoStop}%
\bibitem [{\citenamefont {Zhang}\ \emph
  {et~al.}(2020{\natexlab{b}})\citenamefont {Zhang}, \citenamefont {Tang},
  \citenamefont {Lang}, \citenamefont {Yan},\ and\ \citenamefont
  {Zhu}}]{DWZhang2019}%
  \BibitemOpen
  \bibfield  {author} {\bibinfo {author} {\bibfnamefont {D.-W.}\ \bibnamefont
  {Zhang}}, \bibinfo {author} {\bibfnamefont {L.-Z.}\ \bibnamefont {Tang}},
  \bibinfo {author} {\bibfnamefont {L.-J.}\ \bibnamefont {Lang}}, \bibinfo
  {author} {\bibfnamefont {H.}~\bibnamefont {Yan}}, \ and\ \bibinfo {author}
  {\bibfnamefont {S.-L.}\ \bibnamefont {Zhu}},\ }\href {\doibase
  10.1007/s11433-020-1521-9} {\bibfield  {journal} {\bibinfo  {journal} {Sci.
  China-Phys. Mech. Astron.}\ }\textbf {\bibinfo {volume} {63}},\ \bibinfo
  {pages} {267062} (\bibinfo {year} {2020}{\natexlab{b}})}\BibitemShut
  {NoStop}%
\bibitem [{\citenamefont {Liu}\ \emph {et~al.}(2020{\natexlab{a}})\citenamefont
  {Liu}, \citenamefont {Su}, \citenamefont {Zhang},\ and\ \citenamefont
  {Jiang}}]{Liu_2020}%
  \BibitemOpen
  \bibfield  {author} {\bibinfo {author} {\bibfnamefont {H.-F.}\ \bibnamefont
  {Liu}}, \bibinfo {author} {\bibfnamefont {Z.-X.}\ \bibnamefont {Su}},
  \bibinfo {author} {\bibfnamefont {Z.-Q.}\ \bibnamefont {Zhang}}, \ and\
  \bibinfo {author} {\bibfnamefont {H.}~\bibnamefont {Jiang}},\ }\href
  {\doibase 10.1088/1674-1056/ab8201} {\bibfield  {journal} {\bibinfo
  {journal} {Chin. Phys. B}\ }\textbf {\bibinfo {volume} {29}},\ \bibinfo
  {pages} {050502} (\bibinfo {year} {2020}{\natexlab{a}})}\BibitemShut
  {NoStop}%
\bibitem [{\citenamefont {Zhai}\ \emph {et~al.}(2020)\citenamefont {Zhai},
  \citenamefont {Yin},\ and\ \citenamefont {Huang}}]{ZhaiLiang-Jun2020}%
  \BibitemOpen
  \bibfield  {author} {\bibinfo {author} {\bibfnamefont {L.-J.}\ \bibnamefont
  {Zhai}}, \bibinfo {author} {\bibfnamefont {S.}~\bibnamefont {Yin}}, \ and\
  \bibinfo {author} {\bibfnamefont {G.-Y.}\ \bibnamefont {Huang}},\ }\href
  {\doibase 10.1103/PhysRevB.102.064206} {\bibfield  {journal} {\bibinfo
  {journal} {Phys. Rev. B}\ }\textbf {\bibinfo {volume} {102}},\ \bibinfo
  {pages} {064206} (\bibinfo {year} {2020})}\BibitemShut {NoStop}%
\bibitem [{\citenamefont {Liu}\ \emph {et~al.}(2020{\natexlab{b}})\citenamefont
  {Liu}, \citenamefont {Guo}, \citenamefont {Pu},\ and\ \citenamefont
  {Longhi}}]{LiuTong2020}%
  \BibitemOpen
  \bibfield  {author} {\bibinfo {author} {\bibfnamefont {T.}~\bibnamefont
  {Liu}}, \bibinfo {author} {\bibfnamefont {H.}~\bibnamefont {Guo}}, \bibinfo
  {author} {\bibfnamefont {Y.}~\bibnamefont {Pu}}, \ and\ \bibinfo {author}
  {\bibfnamefont {S.}~\bibnamefont {Longhi}},\ }\href {\doibase
  10.1103/PhysRevB.102.024205} {\bibfield  {journal} {\bibinfo  {journal}
  {Phys. Rev. B}\ }\textbf {\bibinfo {volume} {102}},\ \bibinfo {pages}
  {024205} (\bibinfo {year} {2020}{\natexlab{b}})}\BibitemShut {NoStop}%
\bibitem [{\citenamefont {Hamazaki}\ \emph {et~al.}(2019)\citenamefont
  {Hamazaki}, \citenamefont {Kawabata},\ and\ \citenamefont
  {Ueda}}]{Hamazaki2019}%
  \BibitemOpen
  \bibfield  {author} {\bibinfo {author} {\bibfnamefont {R.}~\bibnamefont
  {Hamazaki}}, \bibinfo {author} {\bibfnamefont {K.}~\bibnamefont {Kawabata}},
  \ and\ \bibinfo {author} {\bibfnamefont {M.}~\bibnamefont {Ueda}},\ }\href
  {\doibase 10.1103/PhysRevLett.123.090603} {\bibfield  {journal} {\bibinfo
  {journal} {Phys. Rev. Lett.}\ }\textbf {\bibinfo {volume} {123}},\ \bibinfo
  {pages} {090603} (\bibinfo {year} {2019})}\BibitemShut {NoStop}%
\bibitem [{\citenamefont {Li}\ \emph {et~al.}(2023)\citenamefont {Li},
  \citenamefont {Ding},\ and\ \citenamefont {Zhang}}]{LiJing2023}%
  \BibitemOpen
  \bibfield  {author} {\bibinfo {author} {\bibfnamefont {J.}~\bibnamefont
  {Li}}, \bibinfo {author} {\bibfnamefont {H.-T.}\ \bibnamefont {Ding}}, \ and\
  \bibinfo {author} {\bibfnamefont {D.-W.}\ \bibnamefont {Zhang}},\ }\href
  {\doibase 10.7498/aps.72.20230862} {\bibfield  {journal} {\bibinfo  {journal}
  {Acta Phys. Sin.}\ }\textbf {\bibinfo {volume} {72}},\ \bibinfo {pages}
  {200601} (\bibinfo {year} {2023})}\BibitemShut {NoStop}%
\bibitem [{\citenamefont {Li}\ \emph {et~al.}(2024)\citenamefont {Li},
  \citenamefont {Yu},\ and\ \citenamefont {Li}}]{ShanZhong2024}%
  \BibitemOpen
  \bibfield  {author} {\bibinfo {author} {\bibfnamefont {S.-Z.}\ \bibnamefont
  {Li}}, \bibinfo {author} {\bibfnamefont {X.-J.}\ \bibnamefont {Yu}}, \ and\
  \bibinfo {author} {\bibfnamefont {Z.}~\bibnamefont {Li}},\ }\href {\doibase
  10.1103/PhysRevB.109.024306} {\bibfield  {journal} {\bibinfo  {journal}
  {Phys. Rev. B}\ }\textbf {\bibinfo {volume} {109}},\ \bibinfo {pages}
  {024306} (\bibinfo {year} {2024})}\BibitemShut {NoStop}%
\bibitem [{\citenamefont {Bergholtz}\ \emph {et~al.}(2021)\citenamefont
  {Bergholtz}, \citenamefont {Budich},\ and\ \citenamefont
  {Kunst}}]{bergholtz2021exceptional}%
  \BibitemOpen
  \bibfield  {author} {\bibinfo {author} {\bibfnamefont {E.~J.}\ \bibnamefont
  {Bergholtz}}, \bibinfo {author} {\bibfnamefont {J.~C.}\ \bibnamefont
  {Budich}}, \ and\ \bibinfo {author} {\bibfnamefont {F.~K.}\ \bibnamefont
  {Kunst}},\ }\href {\doibase 10.1103/RevModPhys.93.015005} {\bibfield
  {journal} {\bibinfo  {journal} {Reviews of Modern Physics}\ }\textbf
  {\bibinfo {volume} {93}},\ \bibinfo {pages} {015005} (\bibinfo {year}
  {2021})}\BibitemShut {NoStop}%
\bibitem [{\citenamefont {Alvarez}\ \emph {et~al.}(2018)\citenamefont
  {Alvarez}, \citenamefont {Vargas},\ and\ \citenamefont
  {Torres}}]{alvarez2018non}%
  \BibitemOpen
  \bibfield  {author} {\bibinfo {author} {\bibfnamefont {V.~M.}\ \bibnamefont
  {Alvarez}}, \bibinfo {author} {\bibfnamefont {J.~B.}\ \bibnamefont {Vargas}},
  \ and\ \bibinfo {author} {\bibfnamefont {L.~F.}\ \bibnamefont {Torres}},\
  }\href {\doibase 10.1103/PhysRevB.97.121401} {\bibfield  {journal} {\bibinfo
  {journal} {Phys. Rev. B}\ }\textbf {\bibinfo {volume} {97}},\ \bibinfo
  {pages} {121401} (\bibinfo {year} {2018})}\BibitemShut {NoStop}%
\bibitem [{\citenamefont {Agarwal}\ \emph {et~al.}({\natexlab{a}})\citenamefont
  {Agarwal}, \citenamefont {Konar}, \citenamefont {Lakkaraju},\ and\
  \citenamefont {De}}]{agarwal2022detecting}%
  \BibitemOpen
  \bibfield  {author} {\bibinfo {author} {\bibfnamefont {K.~D.}\ \bibnamefont
  {Agarwal}}, \bibinfo {author} {\bibfnamefont {T.~K.}\ \bibnamefont {Konar}},
  \bibinfo {author} {\bibfnamefont {L.~G.~C.}\ \bibnamefont {Lakkaraju}}, \
  and\ \bibinfo {author} {\bibfnamefont {A.~S.}\ \bibnamefont {De}},\
  }\href@noop {} {} ({\natexlab{a}}),\ \Eprint
  {http://arxiv.org/abs/2212.12403} {arXiv:2212.12403 [quant-ph]} \BibitemShut
  {NoStop}%
\bibitem [{\citenamefont {Agarwal}\ \emph {et~al.}({\natexlab{b}})\citenamefont
  {Agarwal}, \citenamefont {Konar}, \citenamefont {Lakkaraju},\ and\
  \citenamefont {De}}]{agarwal2023recognizing}%
  \BibitemOpen
  \bibfield  {author} {\bibinfo {author} {\bibfnamefont {K.~D.}\ \bibnamefont
  {Agarwal}}, \bibinfo {author} {\bibfnamefont {T.~K.}\ \bibnamefont {Konar}},
  \bibinfo {author} {\bibfnamefont {L.~G.~C.}\ \bibnamefont {Lakkaraju}}, \
  and\ \bibinfo {author} {\bibfnamefont {A.~S.}\ \bibnamefont {De}},\
  }\href@noop {} {} ({\natexlab{b}}),\ \Eprint
  {http://arxiv.org/abs/2305.08374} {arXiv:2305.08374 [quant-ph]} \BibitemShut
  {NoStop}%
\bibitem [{\citenamefont {Zhu}\ \emph {et~al.}(2021)\citenamefont {Zhu},
  \citenamefont {Zheng}, \citenamefont {Zhu},\ and\ \citenamefont
  {Palumbo}}]{ZhuYanQing2021}%
  \BibitemOpen
  \bibfield  {author} {\bibinfo {author} {\bibfnamefont {Y.-Q.}\ \bibnamefont
  {Zhu}}, \bibinfo {author} {\bibfnamefont {W.}~\bibnamefont {Zheng}}, \bibinfo
  {author} {\bibfnamefont {S.-L.}\ \bibnamefont {Zhu}}, \ and\ \bibinfo
  {author} {\bibfnamefont {G.}~\bibnamefont {Palumbo}},\ }\href {\doibase
  10.1103/PhysRevB.104.205103} {\bibfield  {journal} {\bibinfo  {journal}
  {Phys. Rev. B}\ }\textbf {\bibinfo {volume} {104}},\ \bibinfo {pages}
  {205103} (\bibinfo {year} {2021})}\BibitemShut {NoStop}%
\bibitem [{\citenamefont {Zhang}\ \emph {et~al.}(2019)\citenamefont {Zhang},
  \citenamefont {Wang},\ and\ \citenamefont {Gong}}]{DJzhang2019}%
  \BibitemOpen
  \bibfield  {author} {\bibinfo {author} {\bibfnamefont {D.-J.}\ \bibnamefont
  {Zhang}}, \bibinfo {author} {\bibfnamefont {Q.-h.}\ \bibnamefont {Wang}}, \
  and\ \bibinfo {author} {\bibfnamefont {J.}~\bibnamefont {Gong}},\ }\href
  {\doibase 10.1103/PhysRevA.99.042104} {\bibfield  {journal} {\bibinfo
  {journal} {Phys. Rev. A}\ }\textbf {\bibinfo {volume} {99}},\ \bibinfo
  {pages} {042104} (\bibinfo {year} {2019})}\BibitemShut {NoStop}%
\bibitem [{\citenamefont {Chen~Ye}\ \emph {et~al.}(2024)\citenamefont
  {Chen~Ye}, \citenamefont {Vleeshouwers}, \citenamefont {Heatley},
  \citenamefont {Gritsev},\ and\ \citenamefont {Morais~Smith}}]{ChenYe2024}%
  \BibitemOpen
  \bibfield  {author} {\bibinfo {author} {\bibfnamefont {C.}~\bibnamefont
  {Chen~Ye}}, \bibinfo {author} {\bibfnamefont {W.~L.}\ \bibnamefont
  {Vleeshouwers}}, \bibinfo {author} {\bibfnamefont {S.}~\bibnamefont
  {Heatley}}, \bibinfo {author} {\bibfnamefont {V.}~\bibnamefont {Gritsev}}, \
  and\ \bibinfo {author} {\bibfnamefont {C.}~\bibnamefont {Morais~Smith}},\
  }\href {\doibase 10.1103/PhysRevResearch.6.023202} {\bibfield  {journal}
  {\bibinfo  {journal} {Phys. Rev. Res.}\ }\textbf {\bibinfo {volume} {6}},\
  \bibinfo {pages} {023202} (\bibinfo {year} {2024})}\BibitemShut {NoStop}%
\bibitem [{\citenamefont {Di~Candia}\ \emph {et~al.}(2023)\citenamefont
  {Di~Candia}, \citenamefont {Minganti}, \citenamefont {Petrovnin},
  \citenamefont {Paraoanu},\ and\ \citenamefont {Felicetti}}]{Di2023}%
  \BibitemOpen
  \bibfield  {author} {\bibinfo {author} {\bibfnamefont {R.}~\bibnamefont
  {Di~Candia}}, \bibinfo {author} {\bibfnamefont {F.}~\bibnamefont {Minganti}},
  \bibinfo {author} {\bibfnamefont {K.}~\bibnamefont {Petrovnin}}, \bibinfo
  {author} {\bibfnamefont {G.}~\bibnamefont {Paraoanu}}, \ and\ \bibinfo
  {author} {\bibfnamefont {S.}~\bibnamefont {Felicetti}},\ }\href {\doibase
  10.1038/s41534-023-00690-z} {\bibfield  {journal} {\bibinfo  {journal} {npj
  Quantum Information}\ }\textbf {\bibinfo {volume} {9}},\ \bibinfo {pages}
  {23} (\bibinfo {year} {2023})}\BibitemShut {NoStop}%
\bibitem [{\citenamefont {Liang}\ \emph {et~al.}(2020)\citenamefont {Liang},
  \citenamefont {Su}, \citenamefont {Xiao}, \citenamefont {Che}, \citenamefont
  {Sanders},\ and\ \citenamefont {Wang}}]{Liang2022}%
  \BibitemOpen
  \bibfield  {author} {\bibinfo {author} {\bibfnamefont {H.}~\bibnamefont
  {Liang}}, \bibinfo {author} {\bibfnamefont {Y.}~\bibnamefont {Su}}, \bibinfo
  {author} {\bibfnamefont {X.}~\bibnamefont {Xiao}}, \bibinfo {author}
  {\bibfnamefont {Y.}~\bibnamefont {Che}}, \bibinfo {author} {\bibfnamefont
  {B.~C.}\ \bibnamefont {Sanders}}, \ and\ \bibinfo {author} {\bibfnamefont
  {X.}~\bibnamefont {Wang}},\ }\href {\doibase 10.1103/PhysRevA.102.013722}
  {\bibfield  {journal} {\bibinfo  {journal} {Phys. Rev. A}\ }\textbf {\bibinfo
  {volume} {102}},\ \bibinfo {pages} {013722} (\bibinfo {year}
  {2020})}\BibitemShut {NoStop}%
\bibitem [{\citenamefont {He}\ \emph {et~al.}(2023)\citenamefont {He},
  \citenamefont {Lu}, \citenamefont {Yao}, \citenamefont {Zhu},\ and\
  \citenamefont {Ai}}]{HeWan-Ting2023}%
  \BibitemOpen
  \bibfield  {author} {\bibinfo {author} {\bibfnamefont {W.-T.}\ \bibnamefont
  {He}}, \bibinfo {author} {\bibfnamefont {C.-W.}\ \bibnamefont {Lu}}, \bibinfo
  {author} {\bibfnamefont {Y.-X.}\ \bibnamefont {Yao}}, \bibinfo {author}
  {\bibfnamefont {H.-Y.}\ \bibnamefont {Zhu}}, \ and\ \bibinfo {author}
  {\bibfnamefont {Q.}~\bibnamefont {Ai}},\ }\href {\doibase
  10.1007/s11467-023-1278-2} {\bibfield  {journal} {\bibinfo  {journal} {Front.
  Phys.}\ }\textbf {\bibinfo {volume} {18}},\ \bibinfo {pages} {31304}
  (\bibinfo {year} {2023})}\BibitemShut {NoStop}%
\bibitem [{\citenamefont {Harper}(1955)}]{harper1955single}%
  \BibitemOpen
  \bibfield  {author} {\bibinfo {author} {\bibfnamefont {P.~G.}\ \bibnamefont
  {Harper}},\ }\href {\doibase 10.1088/0370-1298/68/10/304} {\bibfield
  {journal} {\bibinfo  {journal} {Proc. Phys. Soc. Sect. A}\ }\textbf {\bibinfo
  {volume} {68}},\ \bibinfo {pages} {874} (\bibinfo {year} {1955})}\BibitemShut
  {NoStop}%
\bibitem [{\citenamefont {Aubry}\ and\ \citenamefont
  {Andr{\'e}}(1980)}]{aubry1980analyticity}%
  \BibitemOpen
  \bibfield  {author} {\bibinfo {author} {\bibfnamefont {S.}~\bibnamefont
  {Aubry}}\ and\ \bibinfo {author} {\bibfnamefont {G.}~\bibnamefont
  {Andr{\'e}}},\ }\href@noop {} {\bibfield  {journal} {\bibinfo  {journal}
  {Ann. Israel Phys. Soc}\ }\textbf {\bibinfo {volume} {3}},\ \bibinfo {pages}
  {18} (\bibinfo {year} {1980})}\BibitemShut {NoStop}%
\bibitem [{\citenamefont {Ye}\ \emph {et~al.}()\citenamefont {Ye},
  \citenamefont {Vleeshouwers}, \citenamefont {Heatley}, \citenamefont
  {Gritsev},\ and\ \citenamefont {Smith}}]{ye2023quantum}%
  \BibitemOpen
  \bibfield  {author} {\bibinfo {author} {\bibfnamefont {C.-C.}\ \bibnamefont
  {Ye}}, \bibinfo {author} {\bibfnamefont {W.~L.}\ \bibnamefont
  {Vleeshouwers}}, \bibinfo {author} {\bibfnamefont {S.}~\bibnamefont
  {Heatley}}, \bibinfo {author} {\bibfnamefont {V.}~\bibnamefont {Gritsev}}, \
  and\ \bibinfo {author} {\bibfnamefont {C.~M.}\ \bibnamefont {Smith}},\
  }\href@noop {} {}\Eprint {http://arxiv.org/abs/2305.17675} {arXiv:2305.17675}
  \BibitemShut {NoStop}%
\bibitem [{\citenamefont {Sun}\ \emph {et~al.}(2022)\citenamefont {Sun},
  \citenamefont {Tang},\ and\ \citenamefont {Kou}}]{sun2022biorthogonal}%
  \BibitemOpen
  \bibfield  {author} {\bibinfo {author} {\bibfnamefont {G.}~\bibnamefont
  {Sun}}, \bibinfo {author} {\bibfnamefont {J.-C.}\ \bibnamefont {Tang}}, \
  and\ \bibinfo {author} {\bibfnamefont {S.-P.}\ \bibnamefont {Kou}},\ }\href
  {\doibase 10.1007/s11467-021-1126-1} {\bibfield  {journal} {\bibinfo
  {journal} {Frontiers of Physics}\ }\textbf {\bibinfo {volume} {17}},\
  \bibinfo {pages} {1} (\bibinfo {year} {2022})}\BibitemShut {NoStop}%
\bibitem [{\citenamefont {Chang}\ \emph {et~al.}(1997)\citenamefont {Chang},
  \citenamefont {Ikezawa},\ and\ \citenamefont {Kohmoto}}]{Chang1997}%
  \BibitemOpen
  \bibfield  {author} {\bibinfo {author} {\bibfnamefont {I.}~\bibnamefont
  {Chang}}, \bibinfo {author} {\bibfnamefont {K.}~\bibnamefont {Ikezawa}}, \
  and\ \bibinfo {author} {\bibfnamefont {M.}~\bibnamefont {Kohmoto}},\ }\href
  {\doibase 10.1103/PhysRevB.55.12971} {\bibfield  {journal} {\bibinfo
  {journal} {Phys. Rev. B}\ }\textbf {\bibinfo {volume} {55}},\ \bibinfo
  {pages} {12971} (\bibinfo {year} {1997})}\BibitemShut {NoStop}%
\bibitem [{\citenamefont {Liu}\ \emph {et~al.}(2015)\citenamefont {Liu},
  \citenamefont {Ghosh},\ and\ \citenamefont {Chong}}]{FLiu2015}%
  \BibitemOpen
  \bibfield  {author} {\bibinfo {author} {\bibfnamefont {F.}~\bibnamefont
  {Liu}}, \bibinfo {author} {\bibfnamefont {S.}~\bibnamefont {Ghosh}}, \ and\
  \bibinfo {author} {\bibfnamefont {Y.~D.}\ \bibnamefont {Chong}},\ }\href
  {\doibase 10.1103/PhysRevB.91.014108} {\bibfield  {journal} {\bibinfo
  {journal} {Phys. Rev. B}\ }\textbf {\bibinfo {volume} {91}},\ \bibinfo
  {pages} {014108} (\bibinfo {year} {2015})}\BibitemShut {NoStop}%
\bibitem [{\citenamefont {Wang}\ \emph {et~al.}(2021)\citenamefont {Wang},
  \citenamefont {Cheng}, \citenamefont {Liu},\ and\ \citenamefont
  {Yu}}]{YWang2021}%
  \BibitemOpen
  \bibfield  {author} {\bibinfo {author} {\bibfnamefont {Y.}~\bibnamefont
  {Wang}}, \bibinfo {author} {\bibfnamefont {C.}~\bibnamefont {Cheng}},
  \bibinfo {author} {\bibfnamefont {X.-J.}\ \bibnamefont {Liu}}, \ and\
  \bibinfo {author} {\bibfnamefont {D.}~\bibnamefont {Yu}},\ }\href {\doibase
  10.1103/PhysRevLett.126.080602} {\bibfield  {journal} {\bibinfo  {journal}
  {Phys. Rev. Lett.}\ }\textbf {\bibinfo {volume} {126}},\ \bibinfo {pages}
  {080602} (\bibinfo {year} {2021})}\BibitemShut {NoStop}%
\bibitem [{\citenamefont {Li}\ and\ \citenamefont {Li}(2024)}]{SZLi2024}%
  \BibitemOpen
  \bibfield  {author} {\bibinfo {author} {\bibfnamefont {S.-Z.}\ \bibnamefont
  {Li}}\ and\ \bibinfo {author} {\bibfnamefont {Z.}~\bibnamefont {Li}},\ }\href
  {\doibase 10.1103/PhysRevB.110.L041102} {\bibfield  {journal} {\bibinfo
  {journal} {Phys. Rev. B}\ }\textbf {\bibinfo {volume} {110}},\ \bibinfo
  {pages} {L041102} (\bibinfo {year} {2024})}\BibitemShut {NoStop}%
\bibitem [{\citenamefont {Shen}\ \emph {et~al.}(2023)\citenamefont {Shen},
  \citenamefont {Chen}, \citenamefont {Aliyu}, \citenamefont {Qin},
  \citenamefont {Zhong}, \citenamefont {Loh},\ and\ \citenamefont
  {Lee}}]{ruizhe2023}%
  \BibitemOpen
  \bibfield  {author} {\bibinfo {author} {\bibfnamefont {R.}~\bibnamefont
  {Shen}}, \bibinfo {author} {\bibfnamefont {T.}~\bibnamefont {Chen}}, \bibinfo
  {author} {\bibfnamefont {M.~M.}\ \bibnamefont {Aliyu}}, \bibinfo {author}
  {\bibfnamefont {F.}~\bibnamefont {Qin}}, \bibinfo {author} {\bibfnamefont
  {Y.}~\bibnamefont {Zhong}}, \bibinfo {author} {\bibfnamefont
  {H.}~\bibnamefont {Loh}}, \ and\ \bibinfo {author} {\bibfnamefont {C.~H.}\
  \bibnamefont {Lee}},\ }\href {\doibase 10.1103/PhysRevLett.131.080403}
  {\bibfield  {journal} {\bibinfo  {journal} {Phys. Rev. Lett.}\ }\textbf
  {\bibinfo {volume} {131}},\ \bibinfo {pages} {080403} (\bibinfo {year}
  {2023})}\BibitemShut {NoStop}%
\bibitem [{\citenamefont {Zhang}\ \emph {et~al.}(2022)\citenamefont {Zhang},
  \citenamefont {Jiang}, \citenamefont {Deng}, \citenamefont {Wang},
  \citenamefont {Chen}, \citenamefont {Zhang}, \citenamefont {Ren},
  \citenamefont {Dong}, \citenamefont {Xu}, \citenamefont {Gao} \emph
  {et~al.}}]{zhang2022digital}%
  \BibitemOpen
  \bibfield  {author} {\bibinfo {author} {\bibfnamefont {X.}~\bibnamefont
  {Zhang}}, \bibinfo {author} {\bibfnamefont {W.}~\bibnamefont {Jiang}},
  \bibinfo {author} {\bibfnamefont {J.}~\bibnamefont {Deng}}, \bibinfo {author}
  {\bibfnamefont {K.}~\bibnamefont {Wang}}, \bibinfo {author} {\bibfnamefont
  {J.}~\bibnamefont {Chen}}, \bibinfo {author} {\bibfnamefont {P.}~\bibnamefont
  {Zhang}}, \bibinfo {author} {\bibfnamefont {W.}~\bibnamefont {Ren}}, \bibinfo
  {author} {\bibfnamefont {H.}~\bibnamefont {Dong}}, \bibinfo {author}
  {\bibfnamefont {S.}~\bibnamefont {Xu}}, \bibinfo {author} {\bibfnamefont
  {Y.}~\bibnamefont {Gao}},  \emph {et~al.},\ }\href@noop {} {\bibfield
  {journal} {\bibinfo  {journal} {Nature}\ }\textbf {\bibinfo {volume} {607}},\
  \bibinfo {pages} {468} (\bibinfo {year} {2022})}\BibitemShut {NoStop}%
\bibitem [{\citenamefont {Guo}\ \emph {et~al.}(2022)\citenamefont {Guo},
  \citenamefont {Yu}, \citenamefont {Hu},\ and\ \citenamefont
  {Li}}]{GuoZheng-Xin2022}%
  \BibitemOpen
  \bibfield  {author} {\bibinfo {author} {\bibfnamefont {Z.-X.}\ \bibnamefont
  {Guo}}, \bibinfo {author} {\bibfnamefont {X.-J.}\ \bibnamefont {Yu}},
  \bibinfo {author} {\bibfnamefont {X.-D.}\ \bibnamefont {Hu}}, \ and\ \bibinfo
  {author} {\bibfnamefont {Z.}~\bibnamefont {Li}},\ }\href {\doibase
  10.1103/PhysRevA.105.053311} {\bibfield  {journal} {\bibinfo  {journal}
  {Phys. Rev. A}\ }\textbf {\bibinfo {volume} {105}},\ \bibinfo {pages}
  {053311} (\bibinfo {year} {2022})}\BibitemShut {NoStop}%
\bibitem [{\citenamefont {Lee}\ and\ \citenamefont {Chan}(2014)}]{TonyE2014}%
  \BibitemOpen
  \bibfield  {author} {\bibinfo {author} {\bibfnamefont {T.~E.}\ \bibnamefont
  {Lee}}\ and\ \bibinfo {author} {\bibfnamefont {C.-K.}\ \bibnamefont {Chan}},\
  }\href {\doibase 10.1103/PhysRevX.4.041001} {\bibfield  {journal} {\bibinfo
  {journal} {Phys. Rev. X}\ }\textbf {\bibinfo {volume} {4}},\ \bibinfo {pages}
  {041001} (\bibinfo {year} {2014})}\BibitemShut {NoStop}%
\bibitem [{\citenamefont {Wick}(1950)}]{WickG.C.1950}%
  \BibitemOpen
  \bibfield  {author} {\bibinfo {author} {\bibfnamefont {G.~C.}\ \bibnamefont
  {Wick}},\ }\href {\doibase 10.1103/PhysRev.80.268} {\bibfield  {journal}
  {\bibinfo  {journal} {Phys. Rev.}\ }\textbf {\bibinfo {volume} {80}},\
  \bibinfo {pages} {268} (\bibinfo {year} {1950})}\BibitemShut {NoStop}%
\bibitem [{\citenamefont {Zhao}\ \emph {et~al.}(2022)\citenamefont {Zhao},
  \citenamefont {Yi}, \citenamefont {Xue},\ and\ \citenamefont
  {You}}]{Zhaozhuan2022}%
  \BibitemOpen
  \bibfield  {author} {\bibinfo {author} {\bibfnamefont {Z.}~\bibnamefont
  {Zhao}}, \bibinfo {author} {\bibfnamefont {T.-C.}\ \bibnamefont {Yi}},
  \bibinfo {author} {\bibfnamefont {M.}~\bibnamefont {Xue}}, \ and\ \bibinfo
  {author} {\bibfnamefont {W.-L.}\ \bibnamefont {You}},\ }\href {\doibase
  10.1103/PhysRevA.105.063306} {\bibfield  {journal} {\bibinfo  {journal}
  {Phys. Rev. A}\ }\textbf {\bibinfo {volume} {105}},\ \bibinfo {pages}
  {063306} (\bibinfo {year} {2022})}\BibitemShut {NoStop}%
\bibitem [{\citenamefont {Orlov}\ \emph {et~al.}()\citenamefont {Orlov},
  \citenamefont {Shlyapnikov},\ and\ \citenamefont
  {Kurlov}}]{orlov2024adiabatic}%
  \BibitemOpen
  \bibfield  {author} {\bibinfo {author} {\bibfnamefont {P.}~\bibnamefont
  {Orlov}}, \bibinfo {author} {\bibfnamefont {G.~V.}\ \bibnamefont
  {Shlyapnikov}}, \ and\ \bibinfo {author} {\bibfnamefont {D.~V.}\ \bibnamefont
  {Kurlov}},\ }\href@noop {} {}\Eprint {http://arxiv.org/abs/2404.12337}
  {arXiv:2404.12337 [quant-ph]} \BibitemShut {NoStop}%
\bibitem [{\citenamefont {Zeng}\ \emph {et~al.}()\citenamefont {Zeng},
  \citenamefont {Cai}, \citenamefont {Wang},\ and\ \citenamefont
  {Sun}}]{zeng2024fidelity}%
  \BibitemOpen
  \bibfield  {author} {\bibinfo {author} {\bibfnamefont {C.-C.}\ \bibnamefont
  {Zeng}}, \bibinfo {author} {\bibfnamefont {Z.}~\bibnamefont {Cai}}, \bibinfo
  {author} {\bibfnamefont {G.-H.}\ \bibnamefont {Wang}}, \ and\ \bibinfo
  {author} {\bibfnamefont {G.}~\bibnamefont {Sun}},\ }\href@noop {} {}\Eprint
  {http://arxiv.org/abs/2404.16704} {arXiv:2404.16704 [cond-mat.dis-nn]}
  \BibitemShut {NoStop}%
\end{thebibliography}%
	
\end{document}